# Diffusivity Reveals Three Distinct Phases of Interlayer Excitons in $MoSe_2/WSe_2$ Heterobilayers

Jue Wang[1], Qianhui Shi[2], En-Min Shih[2], Lin Zhou[1,3], Wenjing Wu[1], Yusong Bai[1], Daniel A. Rhodes[4], Katayun Barmak[5], James Hone[4], Cory R. Dean[2], X.-Y. Zhu[1,*]

[1] Department of Chemistry, Columbia University, New York, NY 10027, USA.
[2] Department of Physics, Columbia University, New York, NY 10027, USA.
[3] College of Engineering and Applied Sciences, Nanjing University, 210093, P. R. China.
[4] Department of Mechanical Engineering, Columbia University, New York, NY 10027, USA.
[5] Department of Applied Physics and Applied Mathematics, Columbia University, New York, NY 10027, USA.
*email: xyzhu@columbia.edu

## ABSTRACT

**Charge separated interlayer excitons in transition metal dichalcogenide (TMDC) heterobilayers are being explored for moiré exciton lattices and exciton condensates. The presence of permanent dipole moments and the poorly screened Coulomb interaction make many body interactions particularly strong for interlayer excitons. Here we reveal two distinct phase transitions for interlayer excitons in the $MoSe_2/WSe_2$ heterobilayer using time and spatially resolved photoluminescence imaging: from trapped excitons in the moiré-potential to the modestly mobile exciton gas as exciton density increases to $n_{e/h} \sim 10^{11}$ cm$^{-2}$ and from the exciton gas to the highly mobile charge separated electron/hole plasma for $n_{e/h} > 10^{12}$ cm$^{-2}$. The latter is the Mott transition and is confirmed in photoconductivity measurements. These findings set fundamental limits for achieving quantum states of interlayer excitons.**

Heterobilayers of semiconducting transition metal dichalcogenides (TMDCs) are excellent models for the exploration of 2D physics. The type-II band alignment at these heterojunctions leads to the spatial separation of the electron-hole pair upon photo-excitation [1–5], forming long-lived interlayer excitons with permanent dipoles that favor many-body interactions, including exciton condensation [6–9]. The formation of moiré super lattices opens the door to a rich spectrum of phenomena from moiré exciton trapping [10,11], to topological mosaics [12] and highly correlated carrier physics [13–15]. While the inter-exciton interaction promotes many body interactions, their properties also depend strongly on density. Recent spectroscopic experiments have suggested, at low densities, a transition from tapped excitons in moiré potentials [10,11,16,17] to exciton gas and, at high densities, a Mott transition at $n_{Mott}$ = 1.6-3x$10^{12}$ $cm^{-2}$ in $MoSe_2/WSe_2$ between the insulating exciton gas and a charge separated electron/hole plasma [3]. However, little is known quantitively about the density at which the first transition occurs and how properties vary across both phase transitions. Here we determine diffusivity of the interlayer electron/hole system in $MoSe_2/WSe_2$ heterobilayers over four and half orders of magnitude in excitation densities ($n_{e/h} = 10^{9-14}$ $cm^{-2}$). We show that diffusivity increases by similar orders of magnitude with increasing $n_{e/h}$, from $\leq 10^{-3}$ $cm^2s^{-1}$ for moiré trapped excitons, to $10^{-3}$ – $10^{-1}$ $cm^2s^{-1}$ for the exciton gas, and $10^{-1}$ - $10^{1}$ for the electron/hole plasma. The metallic nature of the charge separated electron/hole plasma is also confirmed in photo conductivity measurement.

We fabricate the $MoSe_2/WSe_2$ heterobilayers using the transfer stacking technique from high quality monolayers with low defect density (< $10^{11}$ $cm^{-2}$) [18] and with hexagonal boron nitride (h-BN) encapsulation [19] as detailed elsewhere [3,11]. Intrinsic interlayer exciton lifetimes as long as 200 ns and photo-generated $n_{e/h}$ as high as $10^{14}$ $cm^{-2}$ have been demonstrated [3]. Optical microscope images of the three samples used here are shown in Fig. S1. All measurements are carried out at sample temperatures of 4 K and are completely reversible after repeated continuous wave (CW) or pulsed laser radiation. In photoluminescence (PL) imaging measurements, we use two optically bright heterobilayers with small interlayer twist angles. We calibrate $n_{e/h}$ for the optically bright sample under both pulsed and CW conditions as discussed before [3], Fig. S2, and we use the same calibration to estimate $n_{e/h}$ in the optically dark sample used in photoconductivity measurement. The latter is a reasonable approximation as the dominant exciton species are the momentum indirect excitons with electrons in the Q valleys and holes in the K valleys [3,11,20]. PL come minority bright excitons in equilibrium with the majority dark excitons; lifetimes of the former reflect those of the latter, which is expected to be insensitive to angle-alignment.

To measure the diffusion coefficient ($D$), we photo-excite in a selected region of the heterobilayer and determine the spatial expansion of the interlayer electrons/holes by scanning confocal photoluminescence microscopy in both pulsed and CW modes [4,16,21,22]. We use the PL spectrum at ~1.35 eV from interlayer excitons or electron/hole plasmas [3,11] as a characteristic probe. Fig. 1a shows optical image of one sample used in the imaging experiment (Fig. 1 & 2). The blue dashed outline shows the region of $MoSe_2/WSe_2$ heterobilayer. There is a bubble in the lower middle of the heterobilayer area which shows no PL, Fig. 1b, while surrounding area shows bright PL. Among the $MoSe_2/WSe_2$ heterobilayer samples we fabricated [11], this sample shows a high-level of spatial homogeneity in PL spectra (Fig. S3). Spatially resolved PL from most spots in the bright region, Fig. 1c (left, from spots 1-4 in the PL image map), are characteristic of PL from $MoSe_2/WSe_2$ heterobilayers in most reports to date [2–4,23]. This so-called type-II PL was recently discovered to originate from one-dimensional (1D) strained moiré potentials [11]. The changes in peak position from spot to spot may be attributed to heterogeneity in strain and/or electrostatics and there are also spots that show two or more peaks [11] (see Fig. 2a below). In regions with weak emission, near the edge of the heterobilayer region or the bubble, the PL spectra as represented in Fig. 1c (right, from spots 5-8 in the PL image map). This so-called type-I PL, with one-to-two orders of magnitude lower in intensity than those of type-II, features sharp spectral peaks and originates from un-strained zero-dimension (0D) quantum dot arrays of moiré potentials [11]. In diffusion measurements based on total PL intensity, Fig. 1 and Fig. 2 (below), contribution from type-II PL dominates.

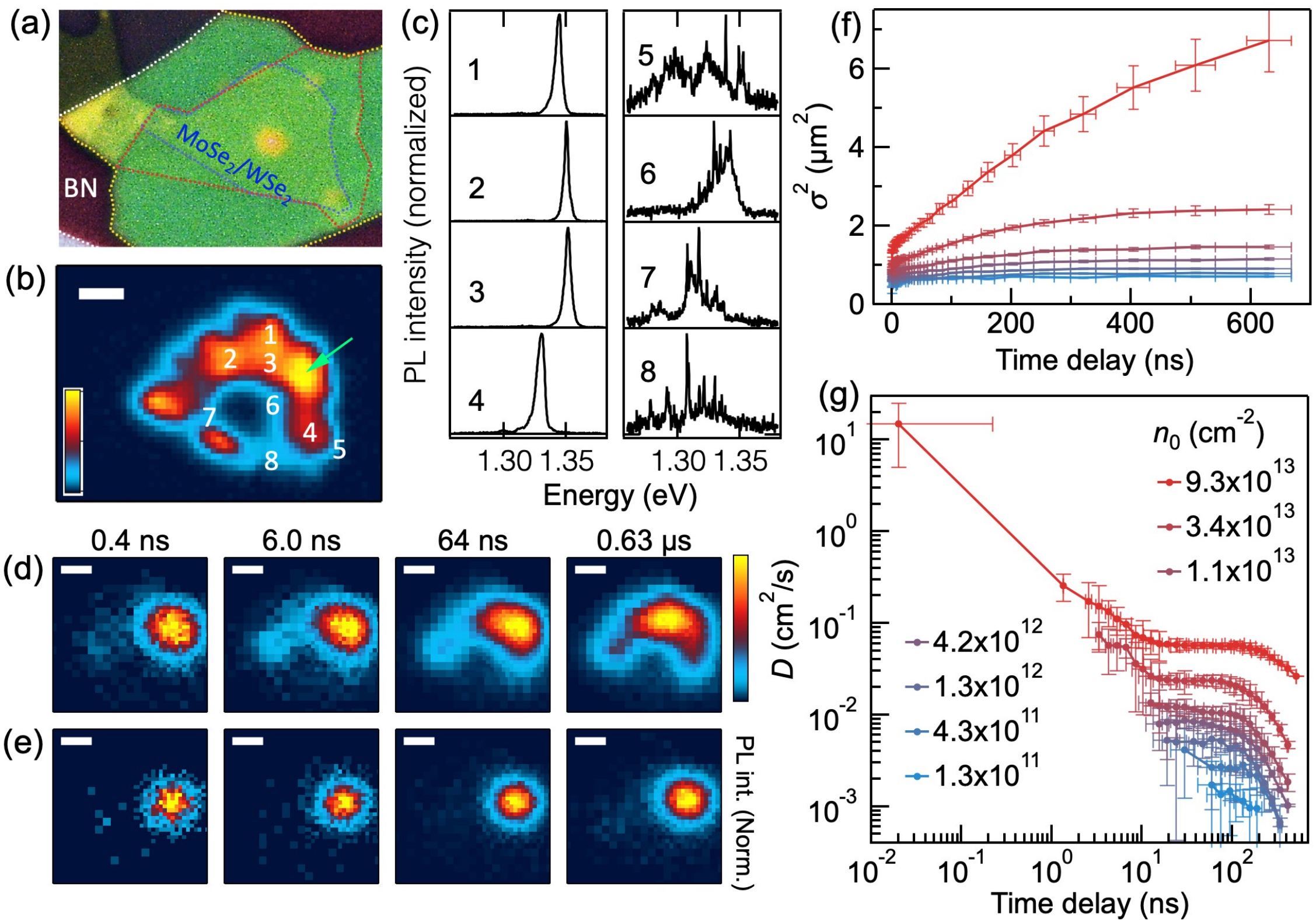

**Fig. 1 | Time-resolved diffusion in $MoSe_2/WSe_2$ heterobilayer. (a)** optical image of the h-BN encapsulated sample. The blue dashed outline shows the $MoSe_2/WSe_2$ heterobilayer region, with a bubble in the lower-middle part. **(b)**, Photoluminescence map from interlayer excitons in the $MoSe_2/WSe_2$ heterobilayer, showing the dark bubble. The numbers (1-8) in **(b)** mark the locations for PL spectra in **(c)**. **(d)** and **(e)** PL images at different $t$ for high ($n_0 = 9.3\times10^{13}$ $cm^{-2}$, **d**) and low ($1.3\times10^{12}$ $cm^{-2}$, **e**) excitation densities. Scale bar, 2 µm. **(f)** Time-dependent spatial variance $\sigma^2$ and (**g**) effective diffusion coefficients as a function of time at different $n_0$. The first data point for $n_0 = 9.3\text{x}10^{13}$ $cm^{-2}$ was estimated from the initial spreading of the image beyond the laser excitation spot. Panels (a) and (c) are reproduced from Supporting Information in ref. [11]. Panels **(b)** and **(c)** were obtained with CW excitation with hν = 2.33 eV, $n_{e/h} = 1.05 \times 10^{10}$ $cm^{-2}$. Panels **(d)-(g)** from pulsed excitation (pulse width = 100 fs) at hν = 2.33 eV. Sample temperature = 4 K.

In the time-resolved experiment, we photo-excite a diffraction-limited spot by an ultrafast laser pulse (width ~100 fs) and determine the PL intensity with temporal (~50 ps) and spatial (~0.2 μm) resolution. Details on the experiments and analysis [24,25] can be found in Supporting Information. Fig. 1d and 1e show PL images at different time delays ($t$) for initial excitation densities of $n_0$ = $9.3\times10^{13}$ $cm^{-2}$ and $n_0$ = $1.3\times10^{12}$ $cm^{-2}$, respectively. Diffusion of the electron/hole is reflected in the spatial expansion of PL image with $t$. The PL image expands faster at a higher $n_0$ than it does at a lower $n_0$. In the former, Fig. 1d, the PL image fills the entire heterobilayer region for $t \geq 0.2$ μs. We fit the radial profile of the PL image at each $t$ to a Gaussian function to obtain the spatial variance $\sigma^2(t)$ at different $n_0$ values, as detailed in Fig. S4 and Fig. S5. Fig. 1f plots $\sigma^2(t)$ as functions of $t$, the slope of which gives the apparent diffusion coefficient, $D(t) = \mathrm{d}\sigma^2(t)/\mathrm{d}t$, Fig. 1g. In the case of high $n_0$, we analyze both homogeneous regions and entire images. The resulting time-dependent variances (Fig. S5) and diffusion constants (Fig. S6) are similar, confirming the reliability of our analysis (Fig. S7).

While $D$ generally decreases with time for each $n_0$, Fig. 1g, the shape of the $D(t)$ curves reveals two distinct transitions: for $n_0 > n_{\mathrm{Mott}}$, $D(t)$ decreases initially until $t \sim 10^1$ ns, beyond which $D(t)$ becomes nearly a plateau; for $t > 10^2$ ns, $D(t)$ decreases again. At each $n_0$, the carrier density is substantially reduced during the initial rapid expansion from the diffraction-limited spot ($\sigma_0 \approx 0.3$ μm) within our time resolution. In the case of $n_0 > n_{\mathrm{Mott}}$, this rapid expansion leads to the Mott transition, the tail of which is captured in the drop of $D(t)$ for $t < 10^1$ ns. This initial expansion may be a combined effect of ballistic transport and repulsive drift [26]. The second step occurs within the exciton phase and is characterized by fast decrease in $D(t)$ for $t > 10^2$ ns and can be attributed to trapping. The formation of the $MoSe_2/WSe_2$ heterobilayer results in a moiré potential landscape consisting of quantum-dot or quantum-wire like local potential wells with depth of the order of 10s-100 meV [27,28]. Trapping of interlayer excitons into the moiré potential wells [10,11] is expected to drastically reduce diffusivity [16], as we explore in more detail below.

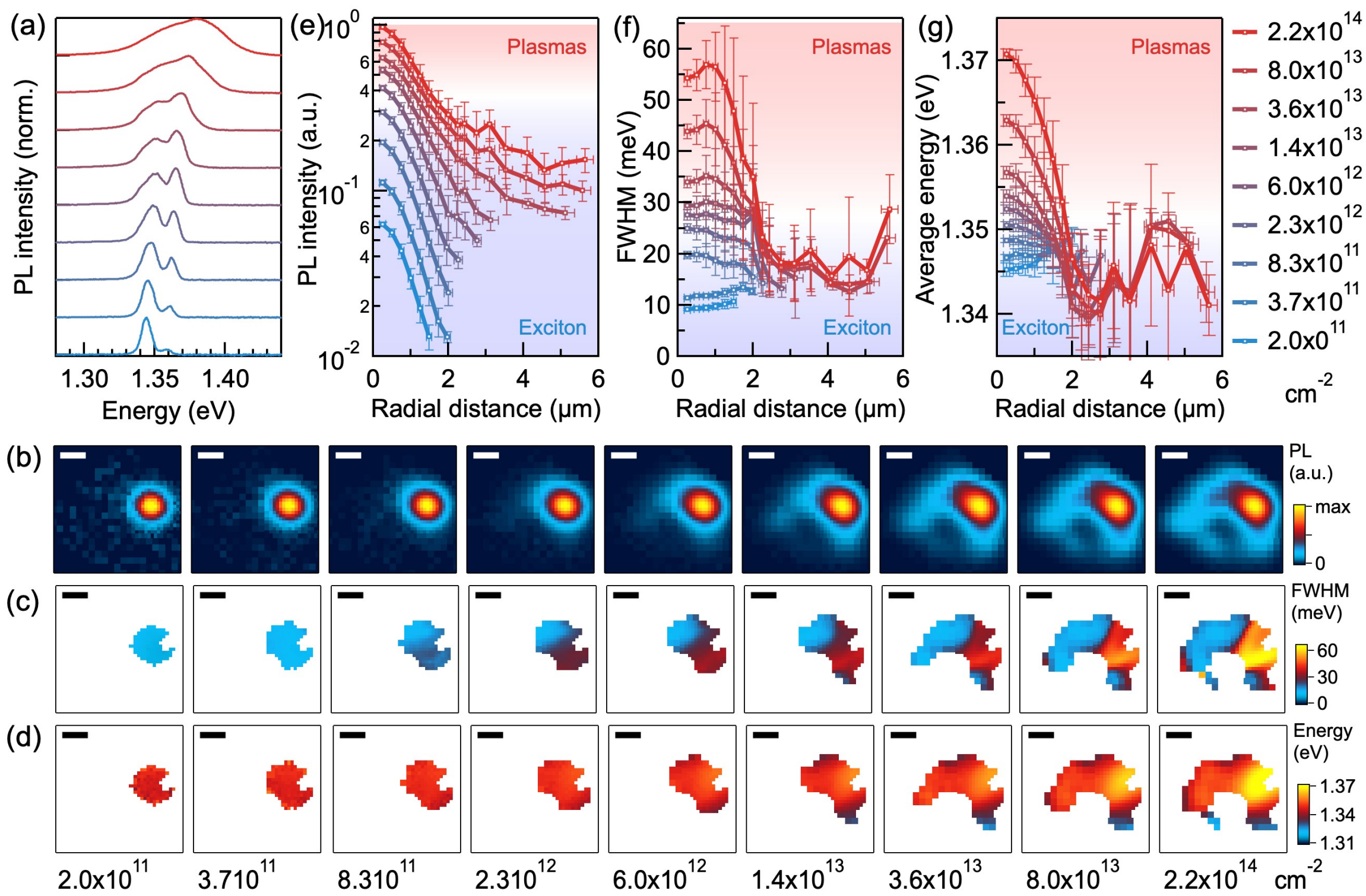

**Fig. 2 | Steady-state diffusion in the $MoSe_2/WSe_2$ heterobilayer.** (**a**) PL spectra at the center excitation spot as a function of excitation density (bottom to top: 2.0x$10^{11}$ – 2.2x$10^{14}$ $cm^{-2}$). Each spectrum is normalized at the peak intensity and offset vertically for clarity). **(b), (c),** and **(d)**, image of PL intensity, FWHM and average energy at different excitation densities, from 2.0×$10^{11}$ $cm^{-2}$ (leftmost) to 2.2×$10^{14}$ $cm^{-2}$ (rightmost), as listed in the legend below **(d)**. Scale bar, 2 μm. (**e**), (**f**), and (**h**), PL intensity, FWHM and average energy as a function of radial distance from excitation spot at different excitation densities. The plasmas phase is within ~ 2 μm of the excitation spot at high $n_{e/h}$. CW excitation at hν = 2.33 eV.

Complementary evidence for the phase transitions can be found in CW PL imaging. In this case, the generation, recombination and diffusion of the interlayer e/h system reach a steady state. For the same sample used in Fig. 1, CW $n_{e/h}$ can reach up to $10^{14}$ $cm^{-2}$ at the center of the excitation spot decays away from the center. The spatial distribution of $n_{e/h}$ allows probing of phases from the local PL spectrum [29], from the dependences on $n_{e/h}$ of the peak intensity ($I_{PL}$), width (full-width-at-half- maximum, FWHM), and average energy ($E_{avg}$). Fig. 2a shows the normalized PL spectra at different densities at the center of the excitation spot (green arrow in Fig. 1b). The spectrum at low $n_{e/h}$ ($10^{11}$ $cm^{-2}$) mainly consists of one peak and, with increasing $n_{e/h}$, a peak on the high energy side grows in relative intensity. These two peaks presumably resulting from two types of strained moiré patterns or dielectric heterogeneity [11]. The peaks broaden with increasing density, particularly across the Mott transition [3] and this process is accompanied by blue shift in average peak energy with increasing carrier density, a result of dipolar repulsion. Fig. S8 summarizes the overall or individual FWHM, and average peak energy, $E_{avg}$ as a function of $n_{e/h}$.

We show spatial maps of $I_{PL}$ (Fig. 2a), FWHM (Fig. 2b), and $E_{avg}$ (Fig. 2c) for a broad range of center excitation densities ($n_c$ = 2.0×10$^{11}$ ~ 2.2×10$^{14}$ cm$^{-2}$). Detailed analysis of the images, including spatial spatially resolved PL spectra at each $n_c$ (Fig. S9) and inhomogeneity-corrected images (Fig. S10), can be found in Supporting Information. For $n_c < n_{Mott}$, both FWHM and $E_{avg}$ are spatially uniform, consistent with a single phase of interlayer exciton gas. At center excitation densities above the $n_{Mott}$, the image maps of both FWHM and $E_{avg}$ clearly show spatial gradients that correspond to the central e/h plasma phase and the exciton gas phase away from the center. We analyze the images quantitatively and show the three quantities as a function of radial distance ($\rho$) from the center, Fig. 2e, 2f, 2g. The plots of FWHM and $E_{avg}$ reveal a phase boundary at $\rho$ ~ 2 μm, within which the system is in the e/h plasma phase at $n_c > n_{Mott}$. For $\rho$ > 2 μm, the system is in the exciton gas phase, independent of $n_c$. From a diffusion length of ~ 2 μm and lifetime of $\tau$ ~ 20 ns, we estimate $D = l^2/\tau$ ~ 2 cm$^2$/s, which is close to the average value for $D$ ~15 – 0.2 cm$^2$/s in the plasma phase from time-resolved imaging experiments. At the high end of CW excitation densities responsible for the plasma phase, the potential presence of thermal gradient may contribute to the diffusion within the $\rho$ ~ 2 μm center region.

To focus on the transition from moiré exciton trapping to an exciton gas, we avoid complications of two types of exciton traps, the unstrained 0D and strained 1D (Fig. 1c). We switch to a $MoSe_2/WSe_2$ heterobilayer sample which is well aligned ($\Delta\theta$ = 1.6°), optically bright, and features only quantum emitter like type-I PL (Fig. S11). Fig. 3a shows $n_{e/h}$-dependent PL spectra from a typical spot on this sample. At $n_{e/h} \leq 10^{11}$ cm$^{-2}$, the multiple sharp PL peaks are indicative of 0D moiré trapped excitons [10,11,17]. For $n_{e/h}$ = ~1-5x10$^{11}$ cm$^{-2}$, the sharp PL peaks merge into one or two broadened peaks, suggesting that the interlayer excitons move out of the 0D moiré traps to form the free exciton gas. Further peak broadening is seen for $n_{e/h} > 10^{12}$ cm$^{-2}$, as expected for the Mott transition. We assign the exciton density of $n_{t-g}$ ~1-5x10$^{11}$ cm$^{-2}$ to the transition from moiré trapped excitons to the free exciton gas due to dipolar repulsion [17]. Fig. 3b plots calculated moiré super cell densities as a function of twist angle (Δθ) for $MoSe_2/WSe_2$ (blue), and five other TMDC heterobilayers. The Δθ = 1.6° for this sample corresponds to a moiré supercell size size of 12 nm and a density of $n_{moiré}$ = 8×10$^{11}$ cm$^{-2}$. The experimental trap-to-gas transition density of $n_{t-g}$

~1-5x10$^{11}$ cm$^{-2}$ is below $n_{moiré}$. Thus, dipolar repulsion makes it improbable to occupy every moiré trap at this twist angle.

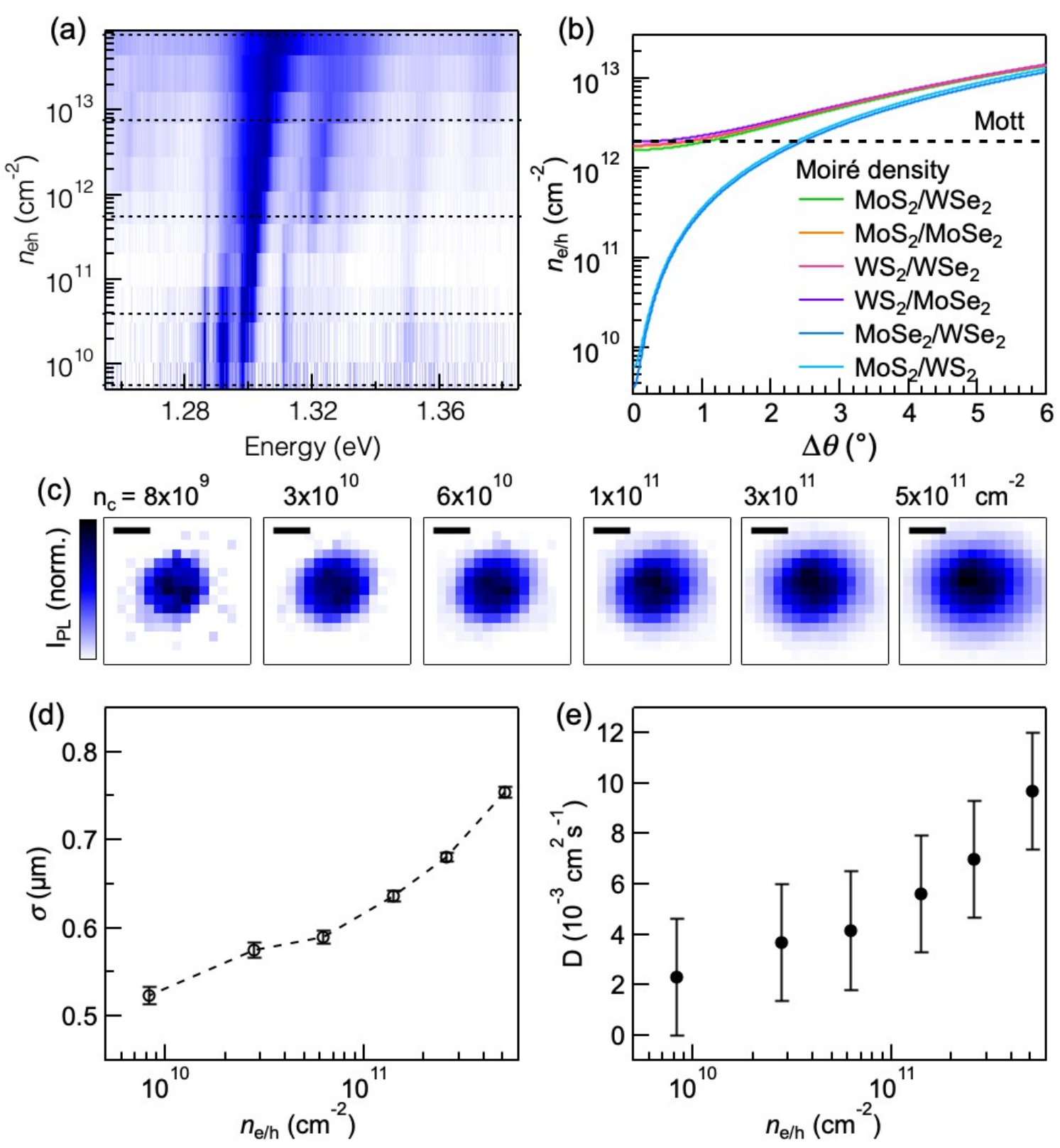

**Fig. 3 | Moiré exciton trapping in the $MoSe_2/WSe_2$ heterobilayer. (a)** Pseudo color (normalized intensity) of PL spectra as a function of excitation density. **(b)** Calculated moiré supercell density as a function of interlayer twist angle for $MoSe_2/WSe_2$ (blue) and five other TMDC heterobilayers. **(c)** CW PL images at different center excitation densities from 8×10$^9$ cm$^{-2}$ to 5×10$^{11}$ cm$^{-2}$. Scale bar, 1 μm. **(d)** Gaussian width at different center densities. **(e)** Apparent diffusion constants estimated from the Gaussian width in (d).

The competition between moiré trapping and mutual exciton repulsion is confirmed in CW PL imaging. We show in Fig. 3c the PL intensity ($I_{PL}$) images for a series of center excitation densities ($n_c$ = 8×10$^9$ - 5×10$^{11}$cm$^{-2}$). At the lowest $n_c$, the Gaussian width $\sigma$ is ~ 0.5 μm (Fig. 4b), close to the theoretical diffraction limit of $\sigma^*$ = 0.21$\lambda$/NA = 0.3 μm [30]. We can estimate upper and lower bounds of $D$ by taking the initial size of the excitation spot as either $\sigma^*$ or the lowest experimental value ($\sigma$ ~ 0.5 μm). Fig. 3e shows the estimated diffusion constant as a function of $n_c$; each data point is an average from these two limits. The diffusion constants are in the range of 10$^{-3}$ cm$^2$/s, similar to the lowest values acquired in the time-resolved experiments at $n_{e/h}$ ~10$^{11}$ cm$^{-2}$ (Fig. 1g). The increasing diffusion constant with exciton density is also confirmed in spatial correlation analysis (Fig. S12): at a low $n_c$, the spectral map is mostly correlated with that at the center excitation spot; at higher $n_c$ increases, the spectra map is more correlated with spectral locally.

The discovery of apparent diffusivity for moiré trapped interlayer excitons is surprising, as the depth of moiré exciton traps has been estimated to be 10s-100 meV [27,28], which makes thermally activated hopping negligible at 4 K. One possible explanaiton is the transient presence of non-equilibrium interlayer excitons. There is 300-400 meV excess energy when interlayer excitons form from intralayer excitons [1–3,20]. Before they completely cooled down and localized to the moiré traps, significant diffusion can occur for the hot interlayer exciton gas. The density-dependent apparent diffusion constant, Fig. 3e, can be attributed to the mutual repulsion in the transient and nonequilibrium dipolar exciton gas.

We now return to the Mott transition, the most direct confirmation of which comes from the conductive nature of the e/h plasma. We measure photoconductivity as a function of $n_{e/h}$ across $n_{Mott}$ in four-terminal geometry (Fig. S13) under CW excitation ($h\nu$ = 2.33 eV). Fig. 4 shows conductivity as a function of $n_{e/h}$ at zero gate voltage; results for non-zero gate voltages are shown in Fig. S14. Conductivity is not measurable at $n_{e/h} \leq n_{Mott}$, as expected from the insulating exciton gas. Above $n_{Mott}$, photoconductivity rises rapidly by more than two orders of magnitude as $n_{e/h}$ increases from $4\times10^{12}$ cm$^{-2}$ to $6\times10^{13}$ cm$^{-2}$. This result provides direct evidence for the insulator-to-metal transition [31]. The threshold $n_{e/h}$ for photoconductivity is in excellent agreement with the spectroscopically or theoretically determined $n_{Mott} \sim 1.6\text{-}3\times10^{12}$ cm$^{-2}$ [3].

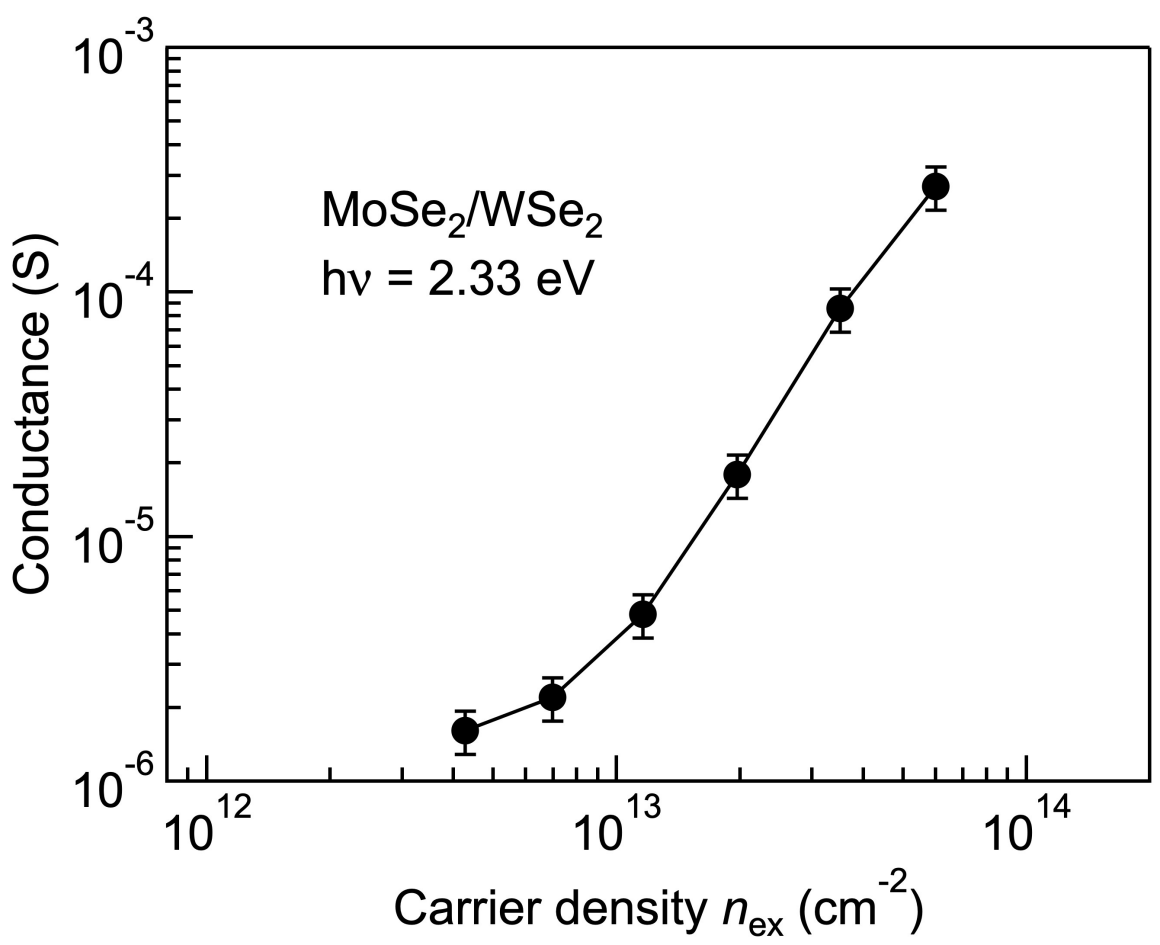

**Fig. 4 | Photoconductance of $MoSe_2/WSe_2$ heterobilayer.** Conductivity as a function of photo-excited density at hν = 2.33 eV showing an insulator state to metal transition. The onset of photoconductivity is close to the Mott density, $n_{Mott}$ ~1.6-3x$10^{12}$ cm$^{-2}$.

Our results quantitatively establish two phase transitions in the $MoSe_2/WSe_2$ heterobilayer: from the moiré trapped excitons to the free interlayer exciton gas and further to the conducting electron/hole plasmas. Interlayer excitons in TMDC heterobilayers are thought to be excellent model systems for exciton condensation [6,7]. Our results indicate that the free interlayer exciton picture, widely assumed in the literature, only exists in the narrow density range around $10^{11} \sim 10^{12}$ cm$^{-2}$. In the low density range, our results also reveal limits for achieving a localized moiré exciton lattice [27] due to dipolar repulsion. At the high $n_{e/h}$ limit, the charge separated nature of the degenerate electron/hole means that the sheet of electrons or holes in each TMDC monolayer is similar to those from gate-doped TMDCs, where superconductivity at 2D carrier densities in the $10^{14}$ $cm^{-2}$ range has been reported [32–35]. This raises the tantalizing possibility of achieving CW photo-induced superconductivity in TMDC heterobilayers, provided cooling of the e/h plasmas can be realized for sufficiently long-lived charge separation.

**Acknowledgements**

We thank Hanqing Xiong for help with scanning confocal microscopy, Kenji Watanabe and Takashi Taniguchi for providing h-BN crystals. All imaging and spectroscopy experiments were supported by the National Science Foundation (NSF) grant DMR-1809680. Sample preparation and photoconductivity measurements were supported in part by the Center for Precision Assembly of Superstratic and Superatomic Solids, a Materials Science and Engineering Research Center (MRSEC) through NSF grant DMR-1420634.

**References**

[1] X. Hong, J. Kim, S. F. Shi, Y. Zhang, C. Jin, Y. Sun, S. Tongay, J. Wu, Y. Zhang, and F. Wang, Nat. Nanotechnol. **9**, 682 (2014).

[2] P. Rivera, J. R. Schaibley, A. M. Jones, J. S. Ross, S. Wu, G. Aivazian, P. Klement, K. Seyler, G. Clark, N. J. Ghimire, J. Yan, D. G. Mandrus, W. Yao, and X. Xu, Nat. Commun. **6**, 6242 (2015).

[3] J. Wang, J. Ardelean, Y. Bai, A. Steinhoff, M. Florian, F. Jahnke, X. Xu, M. Kira, J. Hone, and X.-Y. Zhu, Sci. Adv. **5**, eaax0145 (2019).

[4] L. A. Jauregui, A. Y. Joe, K. Pistunova, D. S. Wild, A. A. High, Y. Zhou, G. Scuri, K. De Greve, A. Sushko, C.-H. Yu, T. Taniguchi, K. Watanabe, D. J. Needleman, M. D. Lukin, H. Park, and P. Kim, Science **366**, 870 (2019).

[5] P. Merkl, F. Mooshammer, P. Steinleitner, A. Girnghuber, K. Q. Lin, P. Nagler, J. Holler, C. Schüller, J. M. Lupton, T. Korn, S. Ovesen, S. Brem, E. Malic, and R. Huber, Nat. Mater. **18**, 691 (2019).

[6] F.-C. Wu, F. Xue, and A. H. MacDonald, Phys. Rev. B **92**, (2015).

[7] M. M. Fogler, L. V Butov, and K. S. Novoselov, Nat. Commun. **5**, 4555 (2014).

[8] O. L. Berman and R. Y. Kezerashvili, Phys. Rev. B **93**, 245410 (2016).

[9] Z. Wang, D. A. Rhodes, K. Watanabe, T. Taniguchi, J. C. Hone, J. Shan, and K. F. Mak, Nature **574**, 76 (2019).

[10] K. L. Seyler, P. Rivera, H. Yu, N. P. Wilson, E. L. Ray, D. G. Mandrus, J. Yan, W. Yao, and X. Xu, Nature **567**, 66 (2019).

[11] Y. Bai, L. Zhou, J. Wang, W. Wu, L. McGilly, D. Halbertal, C. F. B. Lo, F. Liu, J. Ardelean, P. Rivera, N. R. Finney, X. Yang, D. N. Basov, W. Yao, X. Xu, J. Hone, A. Pasupathy, and X.-Y. Zhu, Nat. Mater. **19**, 10.1038/s41563 (2020).

[12] Q. Tong, H. Yu, Q. Zhu, Y. Wang, X. Xu, and W. Yao, Nat. Phys. **13**, 356 (2017).

[13] Y. Tang, L. Li, T. Li, Y. Xu, S. Liu, K. Barmak, K. Watanabe, T. Taniguchi, A. H. MacDonald, J. Shan, and K. F. Mak, Nature **579**, 353 (2020).

[14] E. C. Regan, D. Wang, C. Jin, M. I. Bakti Utama, B. Gao, X. Wei, S. Zhao, W. Zhao, Z. Zhang, K. Yumigeta, M. Blei, J. D. Carlström, K. Watanabe, T. Taniguchi, S. Tongay, M. Crommie, A. Zettl, and F. Wang, Nature **579**, 359 (2020).

[15] Y. Shimazaki, I. Schwartz, K. Watanabe, T. Taniguchi, M. Kroner, and A. Imamoğlu, Nature **580**, 472 (2020).

[16] L. Yuan, B. Zheng, J. Kunstmann, T. Brumme, A. B. Kuc, C. Ma, S. Deng, D. Blach, A. Pan, and L. Huang, Nat. Mater. **19**, 617 (2020).

[17] W. Li, X. Lu, S. Dubey, L. Devenica, and A. Srivastava, Nat. Mater. **19**, 624 (2020).

[18] D. Edelberg, D. Rhodes, A. Kerelsky, B. Kim, J. Wang, A. Zangiabadi, C. Kim, A. Abhinandan, J. Ardelean, M. Scully, D. Scullion, L. Embon, R. Zu, E. J. G. Santos, L. Balicas, C. Marianetti, K. Barmak, X. Zhu, J. Hone, and A. N. Pasupathy, Nano Lett. **19**, 4371 (2019).

[19] O. Ajayi, J. Ardelean, G. Shepard, J. Wang, A. Antony, T. Taniguchi, K. Watanabe, T. Heinz, S. Strauf, and X. Y. Zhu, 2D Mater. (2017).

[20] R. Gillen and J. Maultzsch, Phys. Rev. B **97**, 1 (2018).

[21] P. Rivera, K. L. Seyler, H. Yu, J. R. Schaibley, J. Yan, D. G. Mandrus, W. Yao, and X. Xu, Science (80-. ). **351**, 688 (2016).

[22] D. Unuchek, A. Ciarrocchi, A. Avsar, Z. Sun, K. Watanabe, T. Taniguchi, and A. Kis, Nat. Nanotechnol. **14**, 1104 (2019).

[23] J. Wang, J. Ardelean, Y. Bai, A. Steinhoff, M. Florian, F. Jahnke, X. Xu, M. Kira, J. Hone, and X. Zhu, Sci. Adv. **5**, eaax0145 (2019).

[24] N. S. Ginsberg and W. A. Tisdale, Annu. Rev. Phys. Chem. **70**, https://doi.org/10.1146/annurev (2020).

[25] T. Zhu, J. M. Snaider, L. Yuan, and L. Huang, Annu. Rev. Phys. Chem. **70**, 219 (2019).

[26] Z. Vörös, R. Balili, D. W. Snoke, L. Pfeiffer, and K. West, Phys. Rev. Lett. **94**, 226401 (2005).

[27] W. Yao, X. Xu, G.-B. Liu, J. Tang, and H. Yu, Sci. Adv. **3**, e1701696 (2017).

[28] K. Tran, G. Moody, F. Wu, X. Lu, J. Choi, K. Kim, A. Rai, D. A. Sanchez, J. Quan, A. Singh, J. Embley, A. Zepeda, M. Campbell, T. Autry, T. Taniguchi, K. Watanabe, N. Lu, S. K. Banerjee, K. L. Silverman, S. Kim, E. Tutuc, L. Yang, A. H. MacDonald, and X. Li, Nature **567**, 71 (2019).

[29] M. Stern, V. Garmider, E. Segre, M. Rappaport, V. Umansky, Y. Levinson, and I. Bar-Joseph, Phys. Rev. Lett. **101**, 257402 (2008).

[30] B. Zhang, J. Zerubia, and J. C. Olivo-Marin, in *Appl. Opt.* (2007).

[31] V. M. Asnin and A. A. Rogachev, JETP Lett. **7**, 360 (1968).

[32] L. J. Li, E. C. T. O'Farrell, K. P. Loh, G. Eda, B. Ozyilmaz, and A. H. C. Neto, Nature **529**, 185 (2016).

[33] Y. Saito, Y. Nakamura, M. S. Bahramy, Y. Kohama, J. Ye, Y. Kasahara, Y. Nakagawa, M. Onga, M. Tokunaga, T. Nojima, Y. Yanase, and Y. Iwasa, Nat. Phys. **12**, 144 (2015).

[34] J. M. Lu, O. Zheliuk, I. Leermakers, N. F. Q. Yuan, U. Zeitler, K. T. Law, and J. T. Ye, Science **350**, 1353 (2015).

[35] J. M. Lu, O. Zheliuk, Q. H. Chen, I. Leermakers, N. E. Hussey, U. Zeitler, and J. T. Ye, Proc. Natl. Acad. Sci. U. S. A. **115**, 3551 (2018).

**Supplementary Materials**

Materials and Methods

Figures S1 to S14.

## Supplementary Materials

# Diffusivity Reveals Three Distinct Phases of Interlayer Excitons in $MoSe_2/WSe_2$ Heterobilayers

Jue Wang[1], Qianhui Shi[2], En-Min Shih[2], Lin Zhou[1,3], Wenjing Wu[1], Yusong Bai[1], Daniel A. Rhodes[4], Katayun Barmak[5], James Hone[4], Cory R. Dean[2], X.-Y. Zhu[1,*]

[1] Department of Chemistry, Columbia University, New York, NY 10027, USA.
[2] Department of Physics, Columbia University, New York, NY 10027, USA.
[3] College of Engineering and Applied Sciences, Nanjing University, 210093, P. R. China.
[4] Department of Mechanical Engineering, Columbia University, New York, NY 10027, USA.
[5] Department of Applied Physics and Applied Mathematics, Columbia University, New York, NY 10027, USA.

*email: xyzhu@columbia.edu

## Materials and Methods

**S1.** Device fabrication. The $WSe_2/MoSe_2$ device was built from exfoliated van der Waals materials using a dry transfer method, following three essential steps. First, the hexagonal boron nitride (hBN) and graphite flakes were picked up layer by layer and released on a silicon substrate, to serve as the bottom dielectric and gate. Second, Pt electrodes were evaporated onto the hBN in a Hall bar geometry. Finally, hBN, monolayer $MoSe_2$ and $WSe_2$ were picked up layer by layer and released onto the bottom hBN with pre-patterned Pt electrodes. All transfer processes were assisted by Polypropylene Carbonate (PPC) conformally wrapped on a polydimethylsiloxane (PDMS) particle fixed on a glass slide, with picking up temperature about 40 - 50 °C and releasing temperature about 120 - 130 °C, respectively.

Three h-BN encapsulated $MoSe_2/WSe_2$ heterobilayer samples were used in the experiments. HS1 ($\Delta\theta = 2.6 \pm 0.5^{o}$) and HS2 ($\Delta\theta = 1.6 \pm 0.4^{o}$) are used in diffusion experiments. HS3 is prepared with Pt electrode contacts and used in photoconductivity experiments. HS1 and HS2 have been characterized before with PL and AFM (In ref. [1], see Fig. S10 for HS1 and Fig. S4 for HS2).

**S2.** Photoluminescence imaging measurements. Photoluminescence imaging measurements were performed on a home-built scanning confocal microscope system based on the liquid helium optical cryostat. The sample was at 4 K and under vacuum. The photoluminescence image was Fourier transformed by the objective to the back focal plane. This Fourier plane was then imaged by a 4-f system consisting of a tube lens (Thorlabs TTL200MP) and a scan lens (Thorlabs SL50-CLS2) so that the pivot point was imaged to the center of a dual-axis galvo mirror scanning system. The spatial scanning of the sample was achieved by scanning the angles of the galvo mirrors. The reflected photoluminescence was spatially filtered with a pinhole placed at a conjugate image plane and sent to the detector. In the mapping-type experiments, the laser excitation and photoluminescence detection spot was scanned together with a shared light path; in the diffusion-type experiments, the laser excitation spot was fixed while the photoluminescence was still imaged by scanning.

In the time-resolved measurements, a pulsed laser (2.33 eV, 150 fs) from a visible optical parametric amplifier (Coherent OPA 9450) pumped by a Ti:sapphire regenerative amplifier (Coherent RegA 9050, 250 kHz, 800 nm, 100 fs) was used to inject a spatial and temporal pulse of carriers; a single-photon avalanche photodiode (IDQ ID100-50) and a time-correlated single-photon counting module (bh SPC-130) was used to collect time-resolved photoluminescence traces. The instrument response function has an FWHM of 100 ps resulting in a time resolution of about 20 ps. In the steady-state measurements, a continuous wave laser (2.33 eV) was used; an InGaAs photodiode array (Princeton Instruments PyLoN-IR) was used to collect photoluminescence spectra. The wavelength was calibrated by neon-argon and mercury atomic emission sources (IntelliCal, Princeton Instruments). The intensity was calibrated by a 400 to 1050 nm tungsten halogen lamp (StellarNet SL1-CAL).

There are three approximations in the analysis of time-resolved PL imaging. The first comes from the limited time resolution (~0.25 ns) of our PL imaging technique. Substantial expansion occurs in the first PL image frame at high $n_0$, beyond the initial diffraction-limited excitation spot. The observed initial spatial variance, $\sigma^2(0)$, increases with $n_0$, as shown by the initial values in Fig. 1f. If we take the initial spatial profile at the lowest excitation density as an upper limit of the excitation profile, $\sigma_0^2 < 0.6$ μm$^2$, and the time resolution as an upper limit of expansion time ($\Delta t < 0.25$ ns), we estimate a lower bound of the initial diffusion coefficient ($D_0$) at the highest excitation density ($n_0 \sim 10^{14}$ cm$^{-2}$) to be ~15 cm$^2$/s (first data point in Fig. 1g). The second approximation comes from the fitting of spatial distribution of PL image to Gaussians, with the implicit assumption of a constant diffusion coefficient [2,3]. Since the diffusion coefficient is not a constant, the obtained $D(t)$ beyond the initial time represents an effective average, accounting for offsets of diffusion coefficients in the plateau region in Fig. 1g. The third approximation is the assumption of PL intensity being proportional to $n_{e/h}$. The oscillator strength is known to be constant below $n_{Mott}$, but decreases above $n_{Mott}$ for intralayer excitons [4]. We may assume a similar change in oscillator strength for the interlayer exciton. Due to the initial fast expansion and the limited time resolution, the results shown in Fig. 1 are for $n_{e/h}$ close to $n_{Mott}$ even for the highest $n_0$ measured here. Thus, a nearly constant oscillator strength for the interlayer exciton is a good approximation.

**S3.** <u>Diffusion-reaction equation</u>

The diffusion-reaction equation with a first order decay term

$$\frac{\partial n}{\partial t} = D\nabla^2 n - \frac{n}{\tau}$$

where $n$ is carrier density, $t$ is time, $D$ is diffusion coefficient and $\tau$ is the decay time constant can be solved by Fourier transforming into the momentum space:

$$\frac{\partial \tilde{n}}{\partial t} = -D\boldsymbol{k}^2\tilde{n} - \frac{\tilde{n}}{\tau}$$

where $\tilde{n}(\boldsymbol{k}, t)$, the Fourier transform of $n(\boldsymbol{r}, t)$, is

$$\tilde{n}(\boldsymbol{k}, t) = \tilde{n}(\boldsymbol{k}, 0)\exp\,(-Dt\boldsymbol{k}^2)\exp\,(-\frac{t}{\tau})$$

The inverse Fourier transform then gives

$$n(\boldsymbol{r}, t) = n(\boldsymbol{r}, 0) * \frac{1}{4\pi Dt}\exp\,(-\frac{\boldsymbol{r}^2}{4Dt})\exp\,(-\frac{t}{\tau})$$

where * denotes convolution. In our time-resolved diffusion experiments, the initial carrier density from photoexcitaiton can be approximated by a 2D Gaussian distribution:

$$n(\boldsymbol{r}, 0) = n_0\exp\,(-\frac{\boldsymbol{r}^2}{2\sigma_0^2})$$

where $n_0 = n(\boldsymbol{0}, 0)$ is the peak carrier density at time zero and center of exciation spot, $\sigma_0$ is the initial spatial variance. Therefore the final solution is an expanding and decaying Gaussian distribution

$$n(\boldsymbol{r}, t) = \frac{n_0\sigma_0^2}{\sigma_0^2 + 2Dt}\exp\left(-\frac{\boldsymbol{r}^2}{4Dt + 2\sigma_0^2} - \frac{t}{\tau}\right)$$

where the square of spatial variance grows linearly with time:

$$\sigma^2 = 2Dt + \sigma_0^2$$

Assuming that (1) the species of interest is independent of other species, (2) the transport is purely diffusional, i.e. the ballistic or drift mechanism are negeligible, (3) the diffusion coefficient is a constant over time and space, (4) the decay term is first order, (5) the initial condition is well described by a Gaussian distribution and (6) the experimental probe is proportional to the carrier density, the diffusion coefficient can be extracted from a linear fitting of the $\sigma^2(t)$ curve.

**S4.** Photoconductivity measurements. In photoconductivity measurements, the sample was mounted in a liquid helium optical cryostat (Montana Instruments Fusion/X-Plane) with a 100×, NA 0.75 objective (Zeiss). A continuous wave laser (2.33 eV) was focused to the back aperture of the objective in order to create an evenly distributed light field to excite the whole sample. The conductivity of the sample under photoexcitation was measured in four-probe configuration where the sample was biased by an alternating current (500 nA, 17.7 Hz) and the voltage was measured at the same frequency with a lock-in amplifier (Stanford Research Systems SR830). Keithley 2400 was used to apply DC gate voltage through the graphene back gate.

We carried out conductance measurement using a non-idea geometry due to the different contact resistances of the different metal leads. We tried different combinations and found the configuration shown in Fig. S2 allowed us to carry out reliable measurements with constant current bias through the source and the drain, and constistent phases on the voltage probes. As shown in Fig. 1 in the main text, we detect non-zero conductance only at excitation densities $\geq 3 \times 10^{12}$ cm$^{-2}$. The presence of sufficiently high photocarriers at or above the Mott transition not only provides conducting carriers, but also turns on the contacts. The results shown in Fig. 1 in the main text are obtained from zero gate bias. We have also carried out photoconductance measurements at finite gate biases. Within the range of gate biases, the conductance comes mainly from photodoping and gate-doping only has minor effects.

**S5.** Supplemental data and analysis

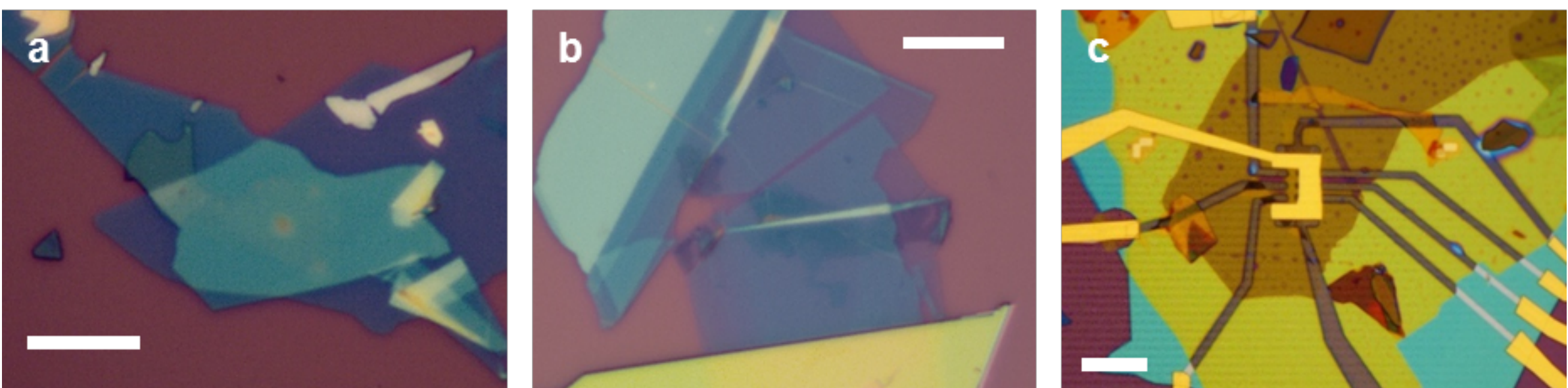

**Fig. S1. $MoSe_2/WSe_2$ heterobilayer samples.** Optical microscope images of thres samples used in the paper. (**a**) HS1 used in diffusion experiments (Fig. 1&2). Twist angle $\Delta\theta = 2.6\pm0.5^{o}$. (**b**) HS2 used in diffusion experiments (Fig. 3). Twist angle $\Delta\theta = 1.6\pm0.4^{o}$. (**c**) HS3 used in photoconductivity experiments (Fig. 4). Detailed characterizations of HS1 in (a) and HS2 in (b) can be found Fig. S10 and Fig. S4, respectively, in ref. [1].

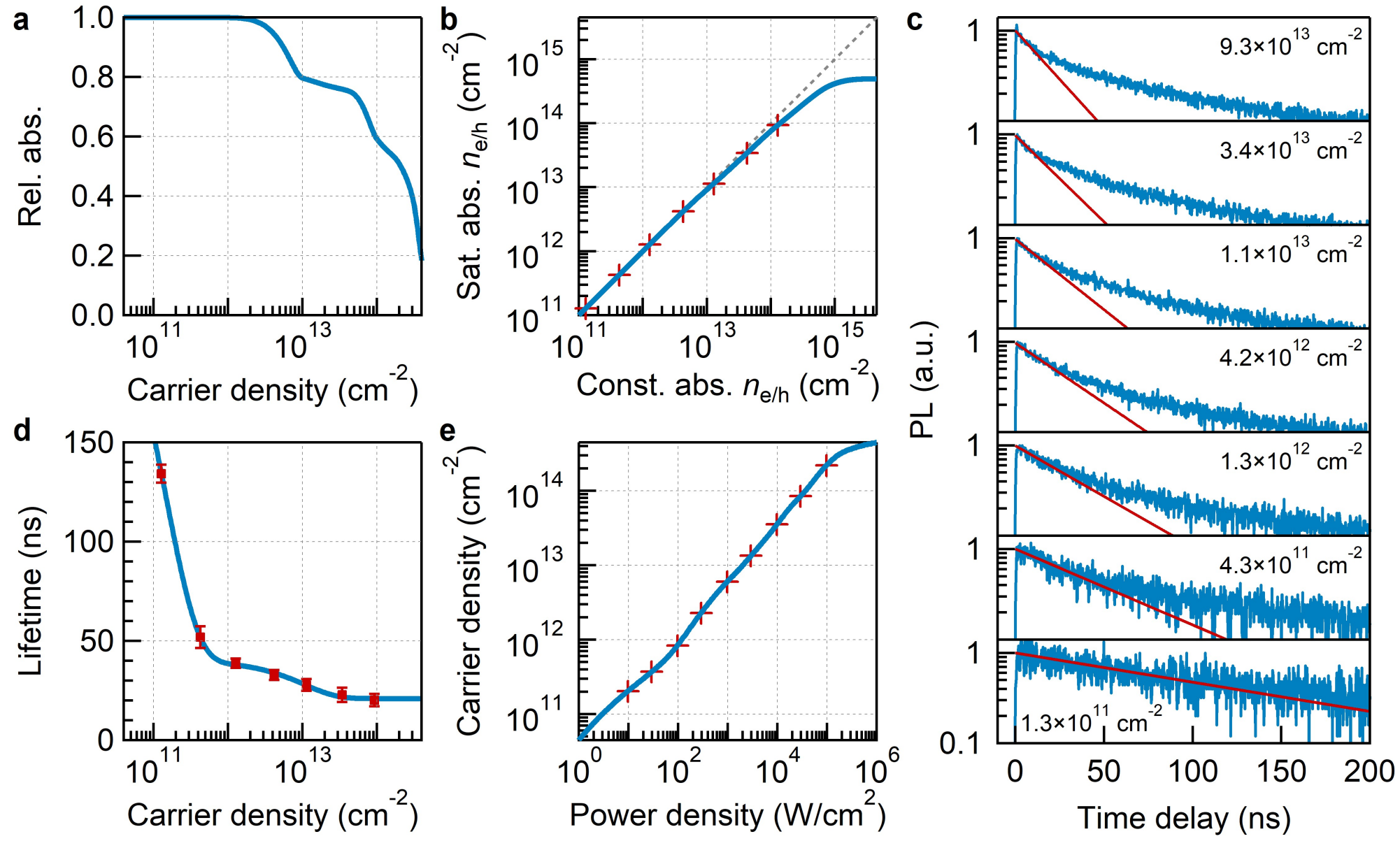

**Fig. S2. Calibration of carrier density in $MoSe_2/WSe_2$ heterobilayers. a**, Relative absorptance as a function of carrier density showing saturable absorption effect, as obtained from theory [4]. **b**, Calibration of carrier density for pulsed excitation. The actual carrier density considering saturable absorption is plotted against carrier density calculated assuming constant absorption. Red plus sign, the carrier density used in time resolved diffusion experiments. **c**, Time-resolved PL at excitation spot ($r < 0.5$ μm) at different excitation densities. The initial decay was fit by single exponential function shown as red lines. **d**, PL lifetime as a function of carrier density (red squares) and apparent bi-exponential fit (blue lines). **e**, Calibration of carrier density for CW excitation. The steady-state carrier density is plotted against excitation power density. Red plus sign, the carrier density at excitation spot used in steady-state diffusion experiments. The laser photon energies for pulsed and steady-state experiments are both 2.33 eV (532 nm).

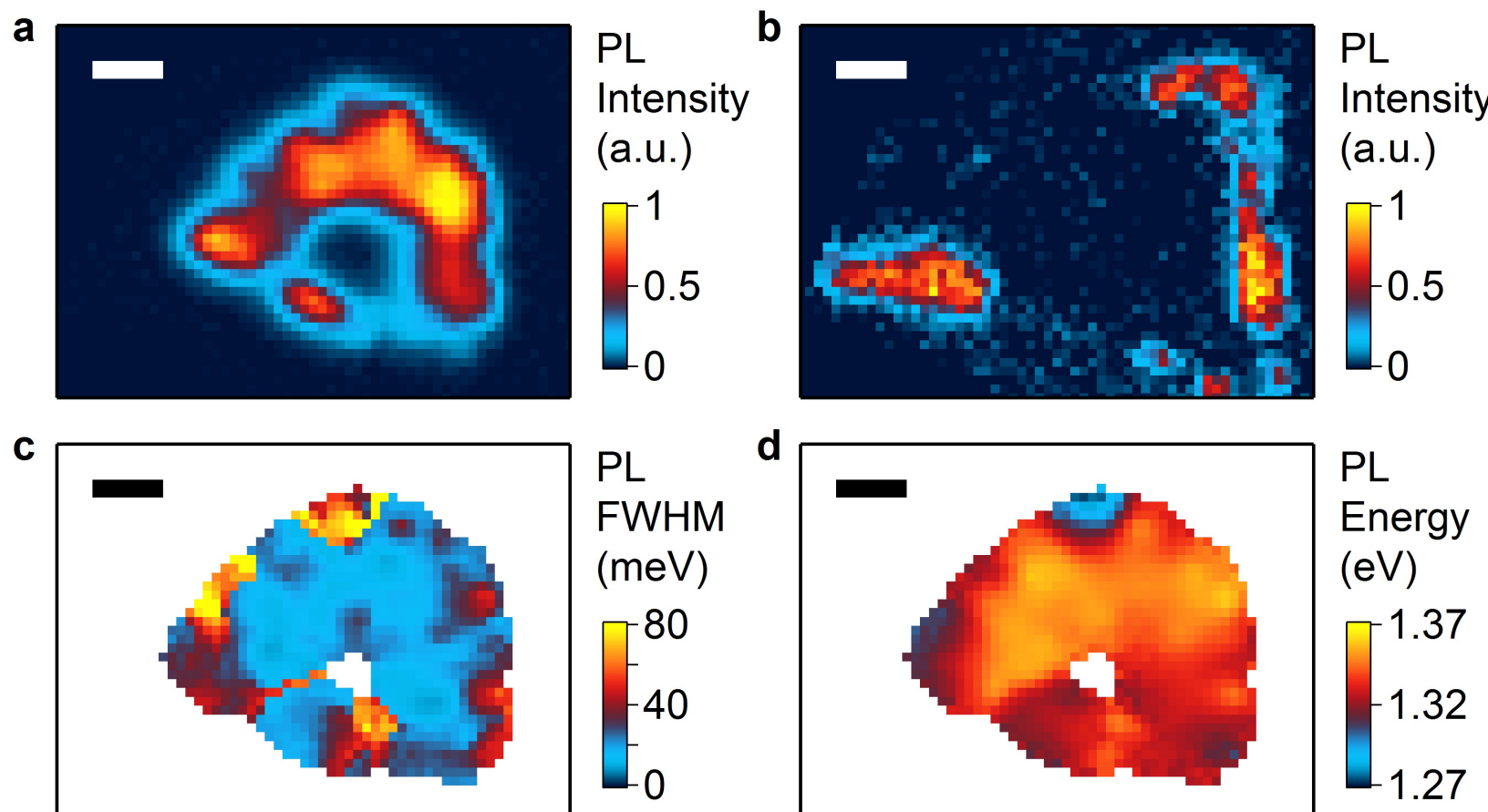

**Fig. S3. Photoluminescence image of the $MoSe_2/WSe_2$ heterobilayer HS1 used in diffusion experiments. a**, Image of interlayer exciton PL intensity. **b**, Image of intralayer exciton PL intensity. **c**, Image of FWHM of interlayer exciton PL. **d**, Image of average energy of interlayer exciton PL. Scale bar, 2 μm.

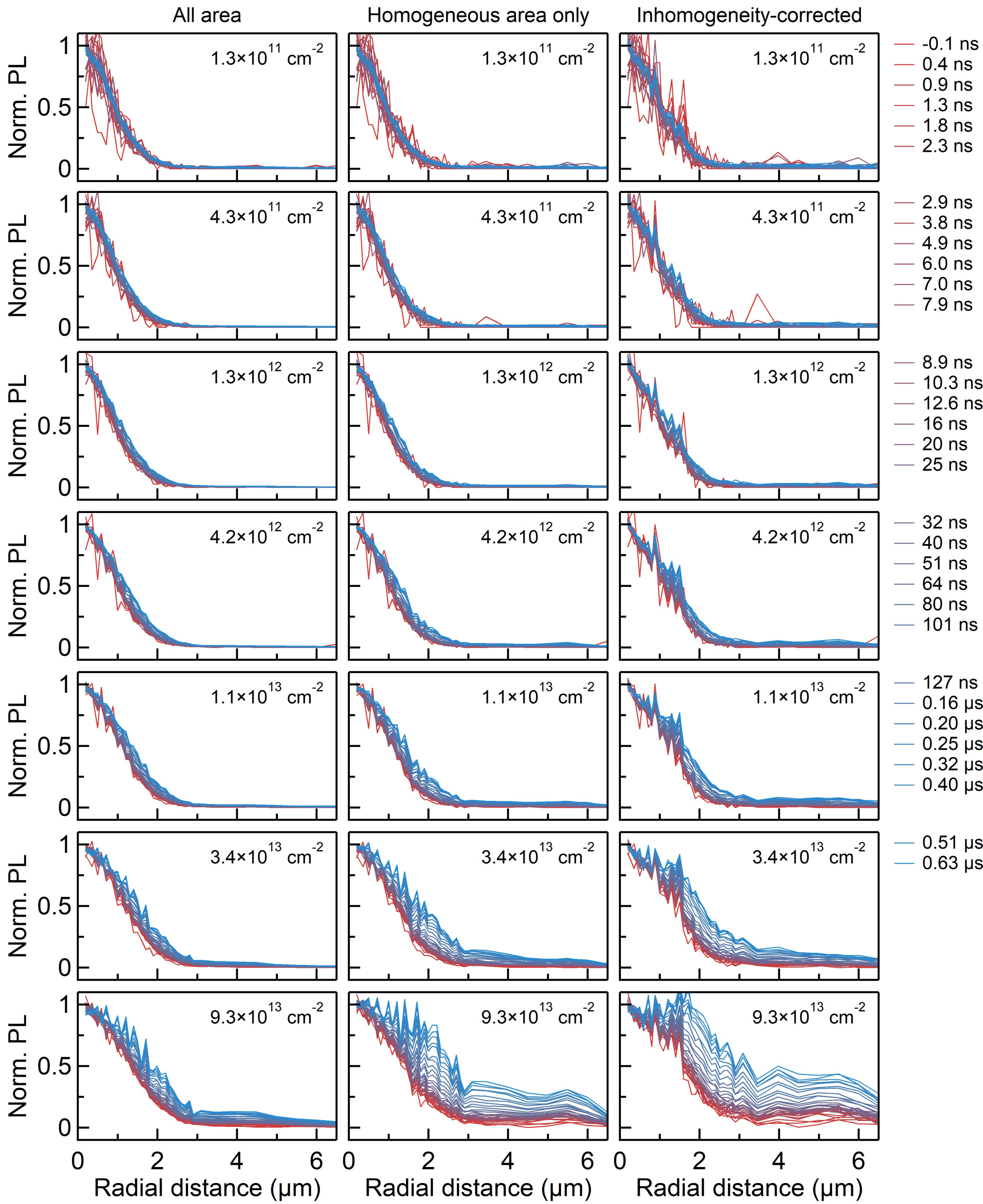

**Fig. S4. Evolution of radial PL profile over time.** Normalized PL as a function of radial distance and delay time (-0.1 ns to 0.63 µs from red to blue lines) for different excitation densities ($1.3\times10^{11}$ to $9.3\times10^{13}$ cm$^{-2}$ from top to bottom) extracted from time-resolved PL imaging data by three methods (from all area, from homogeneous area only and from inhomogeneity-corrected data, from left to right).

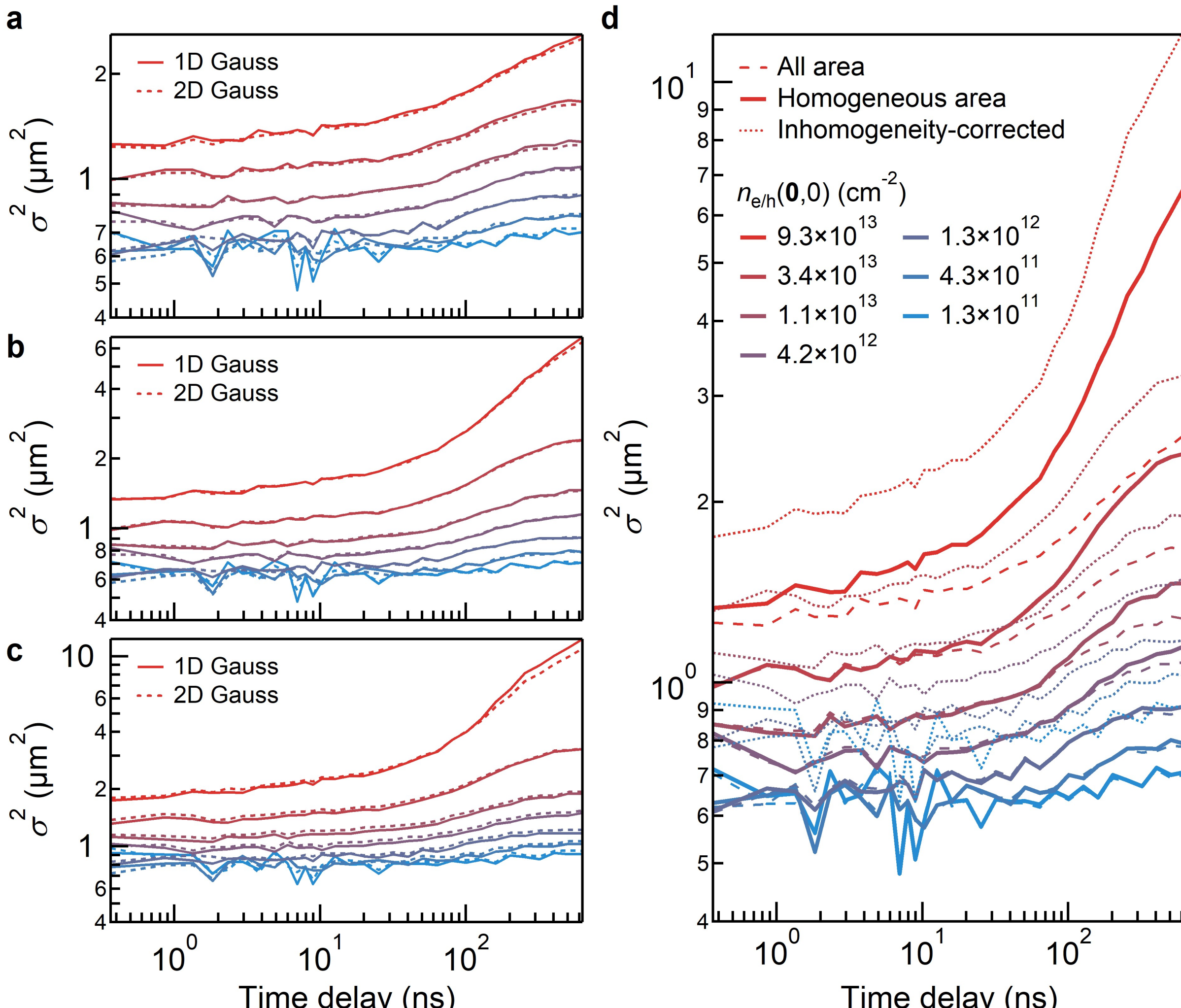

**Fig. S5. Comparison of widths extracted by different methods. a**, Comparison between widths extracted from 2D Gaussian fit to the PL image (dash lines) and from 1D Gaussian fit to the radial PL profile (solid lines), using data from all sample area. **b**, Comparison between 1D and 2D Gaussian widths extracted using data from homogeneous area only. **c**, Comparison between 1D and 2D Gaussian widths extracted from inhomogeneity-corrected data. **d**, Comparison of 1D Gaussian widths extracted from all area, homogeneous area only and inhomogeneity-corrected data. Colors from blue to red indicate different excitation densities $n_{e/h}(\mathbf{0},0)$.

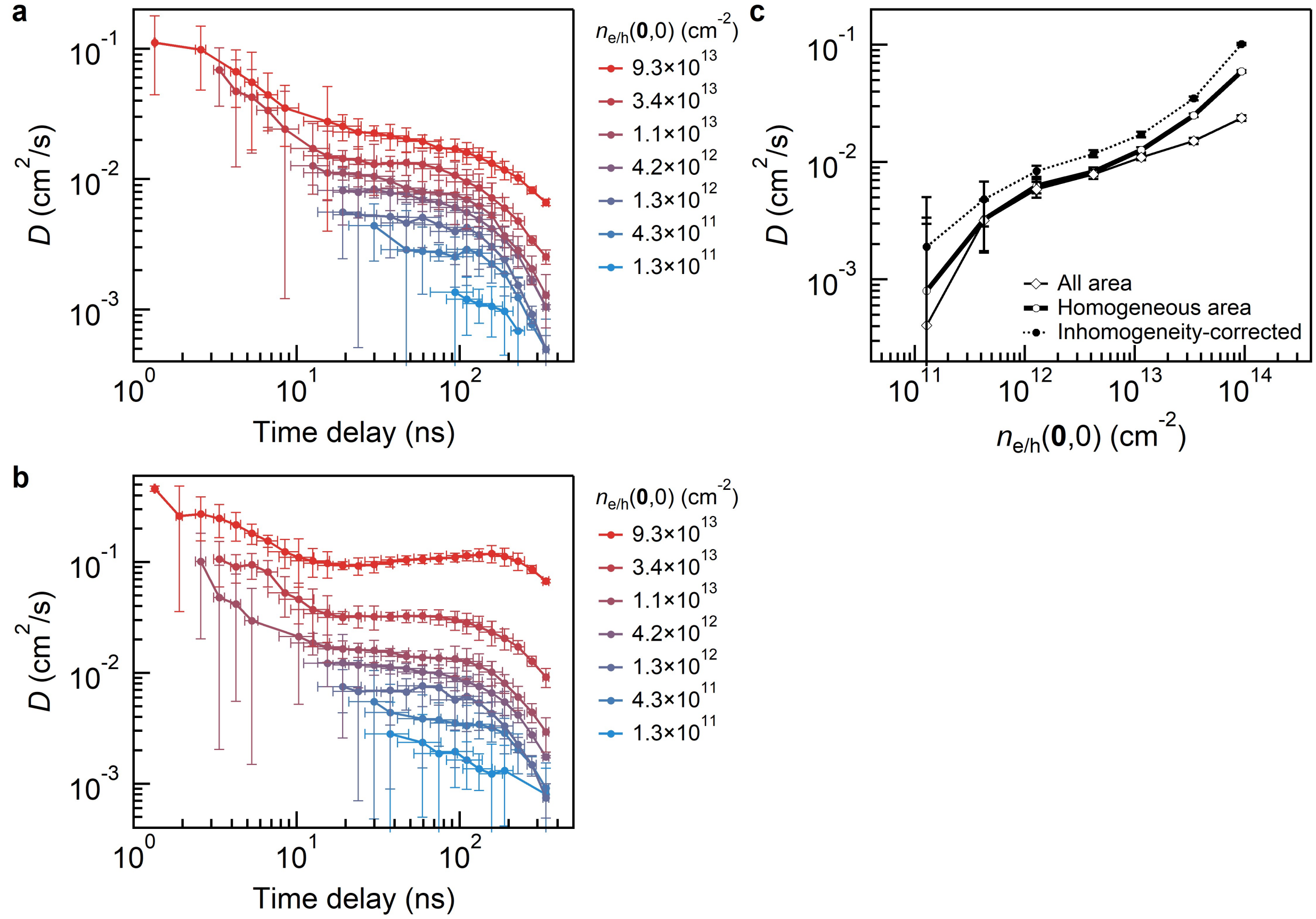

**Fig. S6. Comparison of diffusion coefficients extracted by different methods. a**, Effective diffusion coefficient as a function of time delay and excitation density extracted from image data of all area. **b**, Effective diffusion coefficients extracted from inhomogeneity-corrected image data. **c**, Comparison of plateau diffusion coefficient as a function of excitation density extracted by three methods (from all area, from homogeneous area only and from inhomogeneity-corrected data).

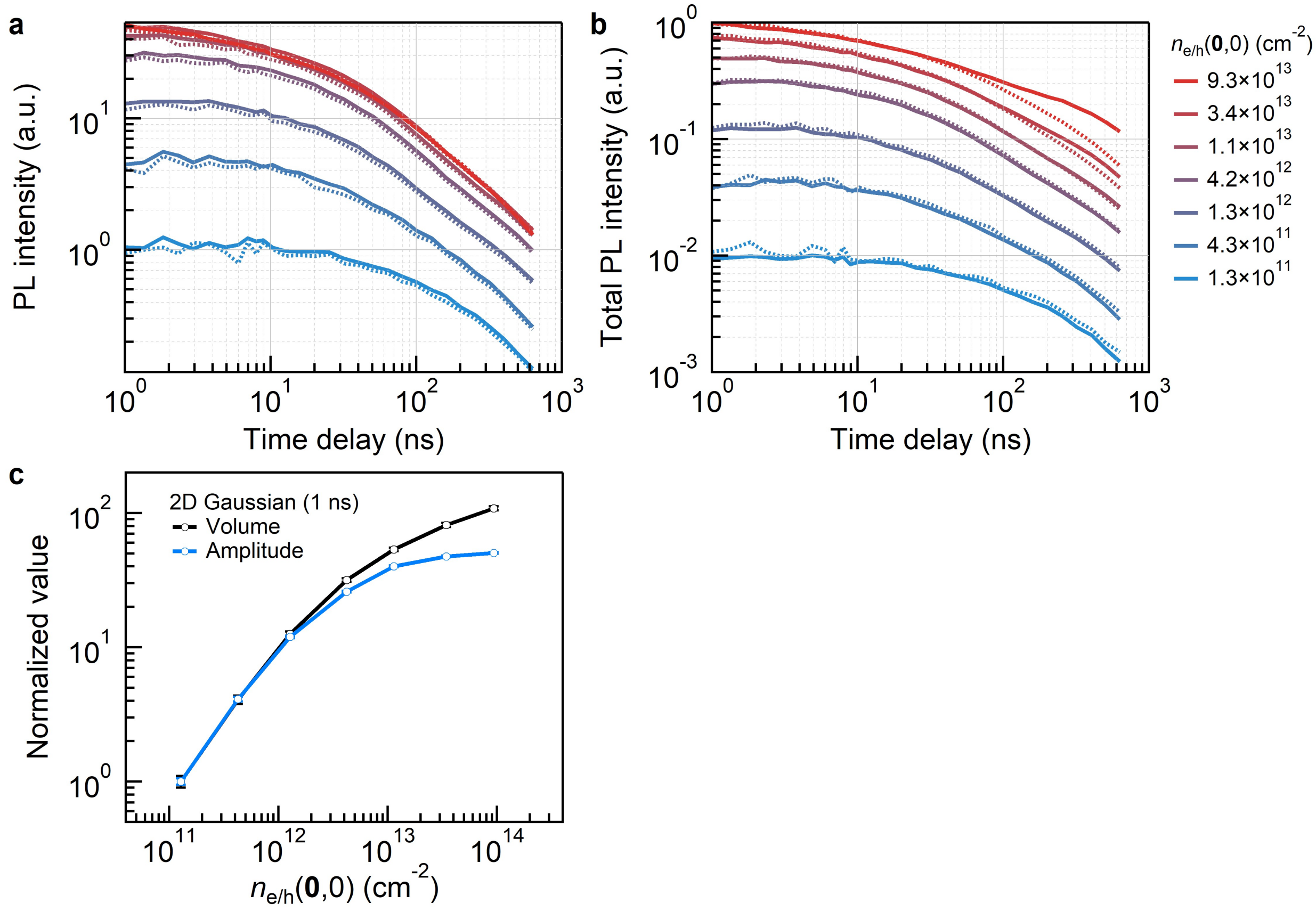

**Fig. S7. Photoluminescence decay. a**, Time resolved PL at center spot ($r < 0.5$ μm, dash lines) or amplitude of 2D Gaussian fit to time resolved PL images (solid lines) at different excitation densities. **b**, Time resolved PL from whole sample (dash lines) or volume of 2D Gaussian fit to time resolved PL images (solid lines) at different excitation densities. **c**, Comparison of excitation density dependence of amplitude and volume of 2D Gaussian fit to PL image at 1 ns delay time.

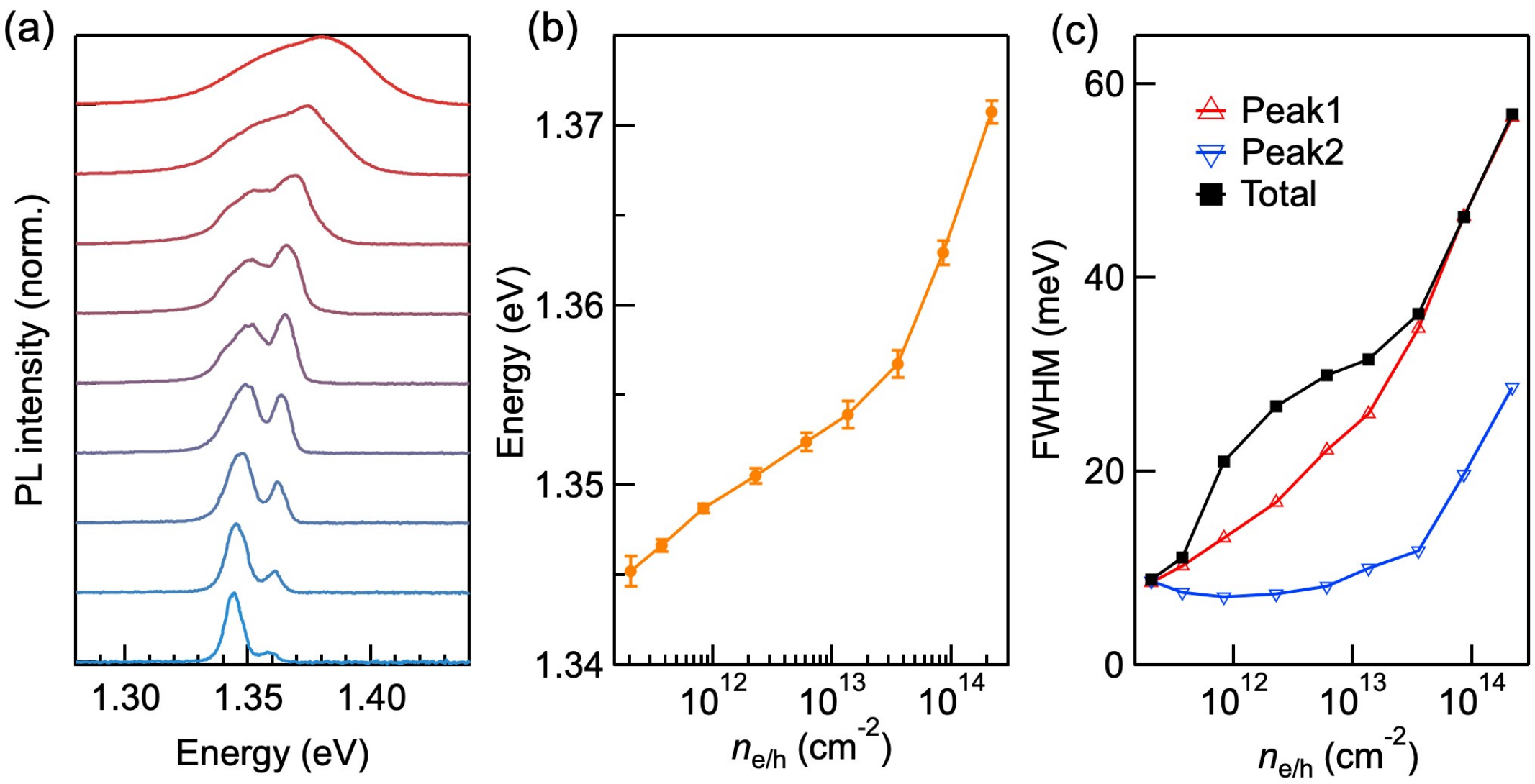

**Fig. S8. PL characteristics across Mott transition. (a)** PL spectra at the center excitation spot as a function of excitation density (bottom to top: 2.0x10$^{11}$ – 2.2x10$^{14}$ cm$^{-2}$) – reproduced from Fig. 2a. (**b**) average energy as a function of excitation density from the spectra in (a). **(c)** Overall FWHM from intensity-weighted peak analysis (black) and individual peak widths ( red and blue) from peak deconvolution with two Gaussians.

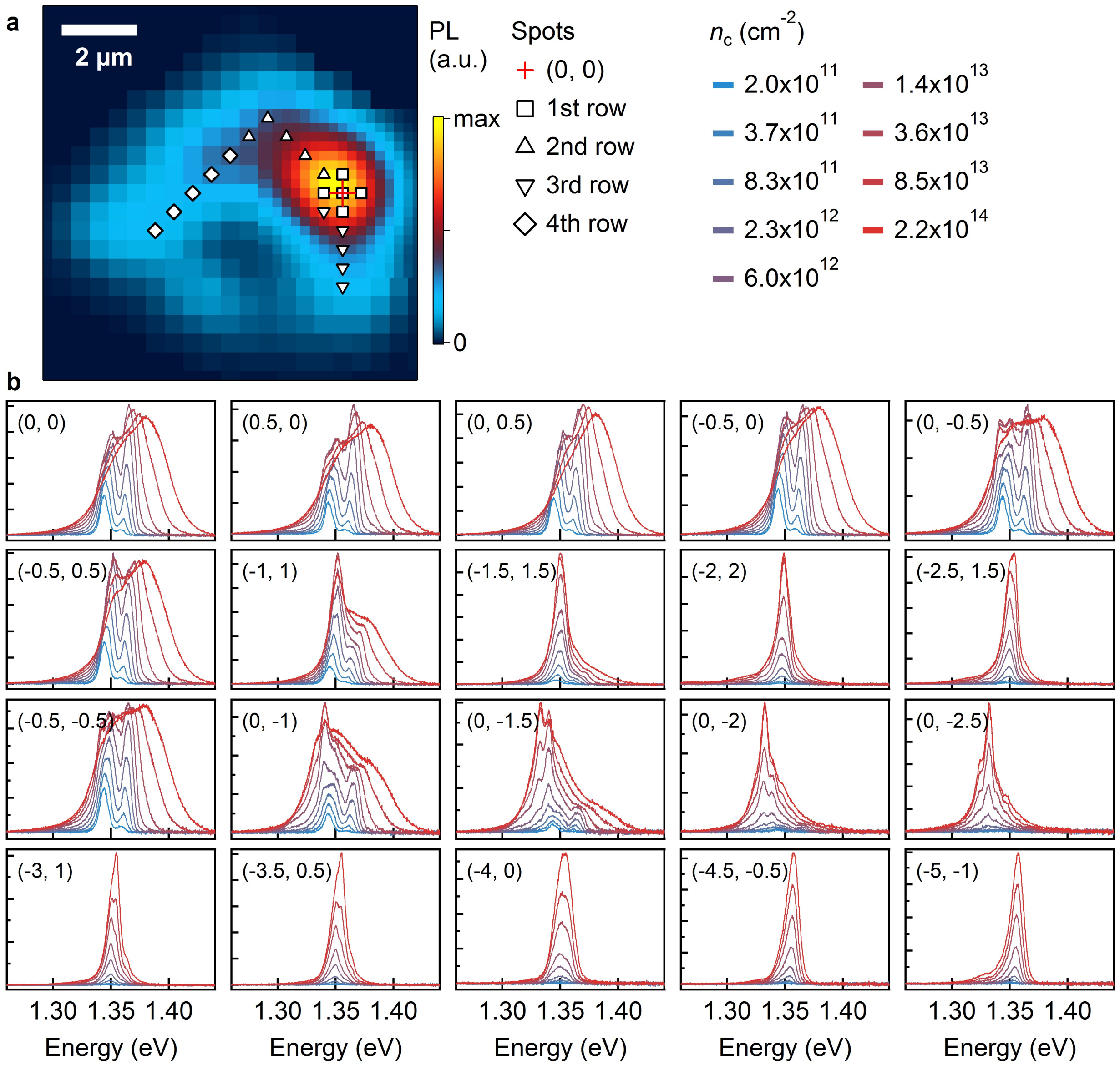

**Fig. S9. Representative PL spectra in the steady-state diffusion experiments.** (**a**) PL intensity image at center excitation density of 2.2×10$^{14}$ cm$^{-2}$, showing the positions of spots selected. (**b**) PL spectra at selected spots at different center excitation densities.

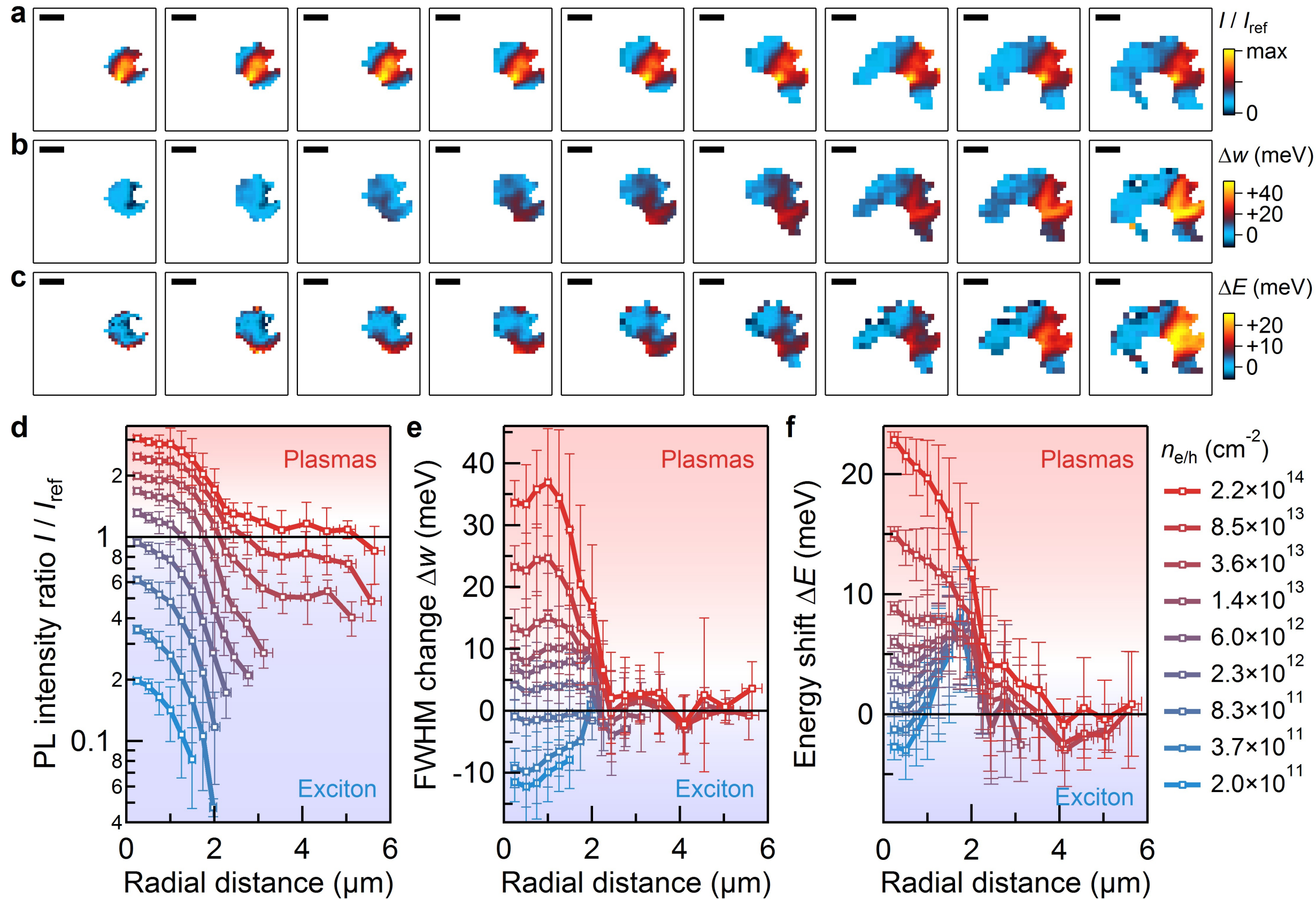

**Fig. S10. Inhomogeneity-corrected PL landscape of steady-state diffusion.** Similar to Fig. 2 but extracted from inhomogeneity-corrected PL image data. **a**, **b**, **c**, The images of PL intensity ratio (**a**), FWHM change (**b**) and average energy shift (**c**) relative to local PL characteristics. The steady-state carrier densities at the excitation spot are $2.0\times10^{11}$, $3.7\times10^{11}$, $8.3\times10^{11}$, $2.3\times10^{12}$, $6.0\times10^{12}$, $1.4\times10^{13}$, $3.6\times10^{13}$, $8.5\times10^{13}$ and $2.2\times10^{14}$ cm$^{-2}$, from left to right. **d**, **e**, **f**, PL intensity ratio (**d**), FWHM change (**e**) and average energy shift (**f**) as a function of radial distance.

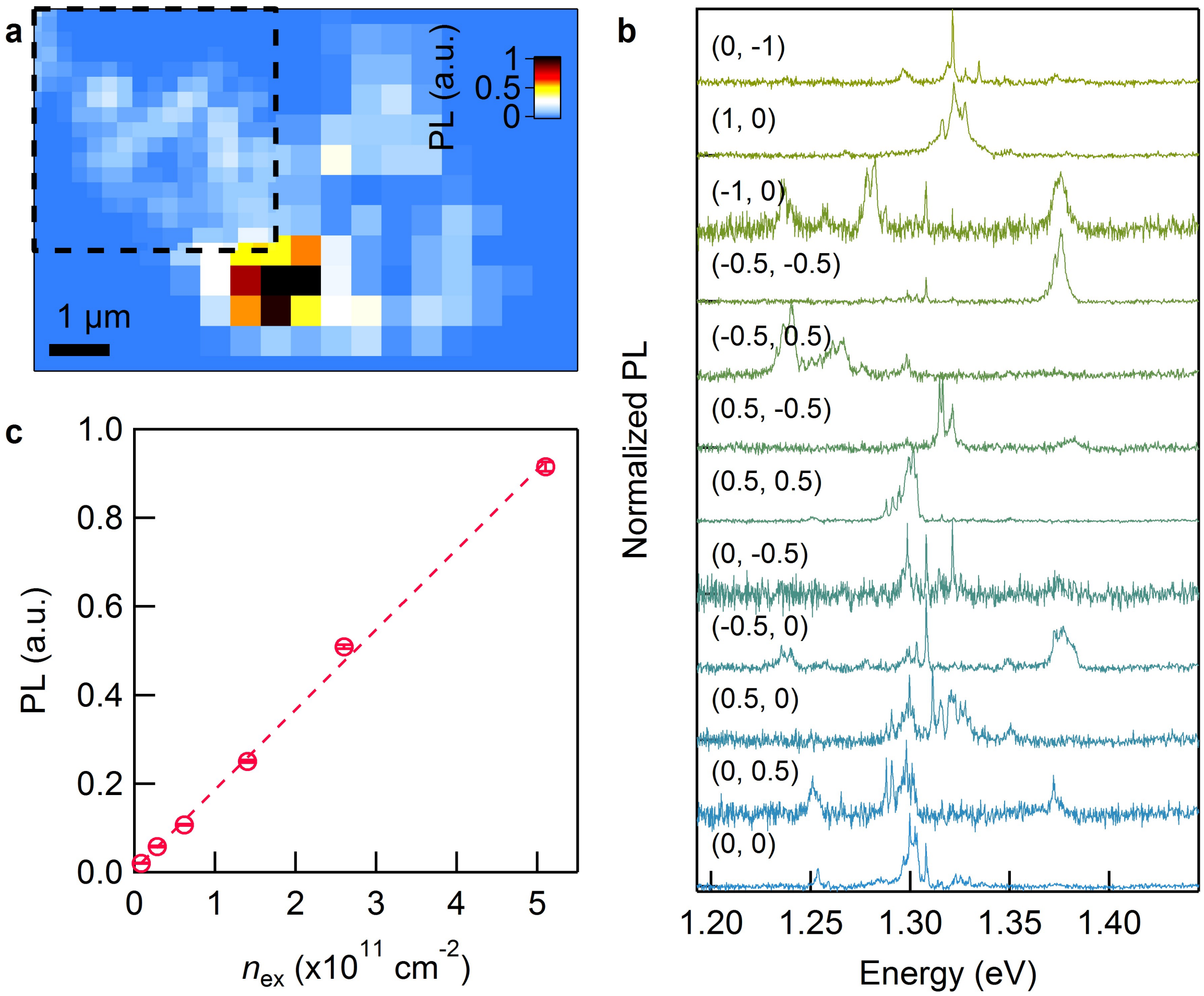

**Fig. S11. PL imaging of HS2.** (**a**) PL mapping of HS2 at about $10^{11}$ $cm^{-2}$ excitation density. The dashed box is relatively homogeneous and is used in the diffusion experiments in Fig. 3. (**b**) PL spectra at selected spot positions. The origin of coordinates is chosen at the center of the dashed box in (**a**). The unit of coordinates is µm. (**c**) PL intensity as a function of excitation density. At this low density regime, PL is linear with power, consistent with the exciton emission mechanism. Other characterizations, including circularly polarized PL from circularly polarized excitation, can be found in Fig. S4 in ref. [1].

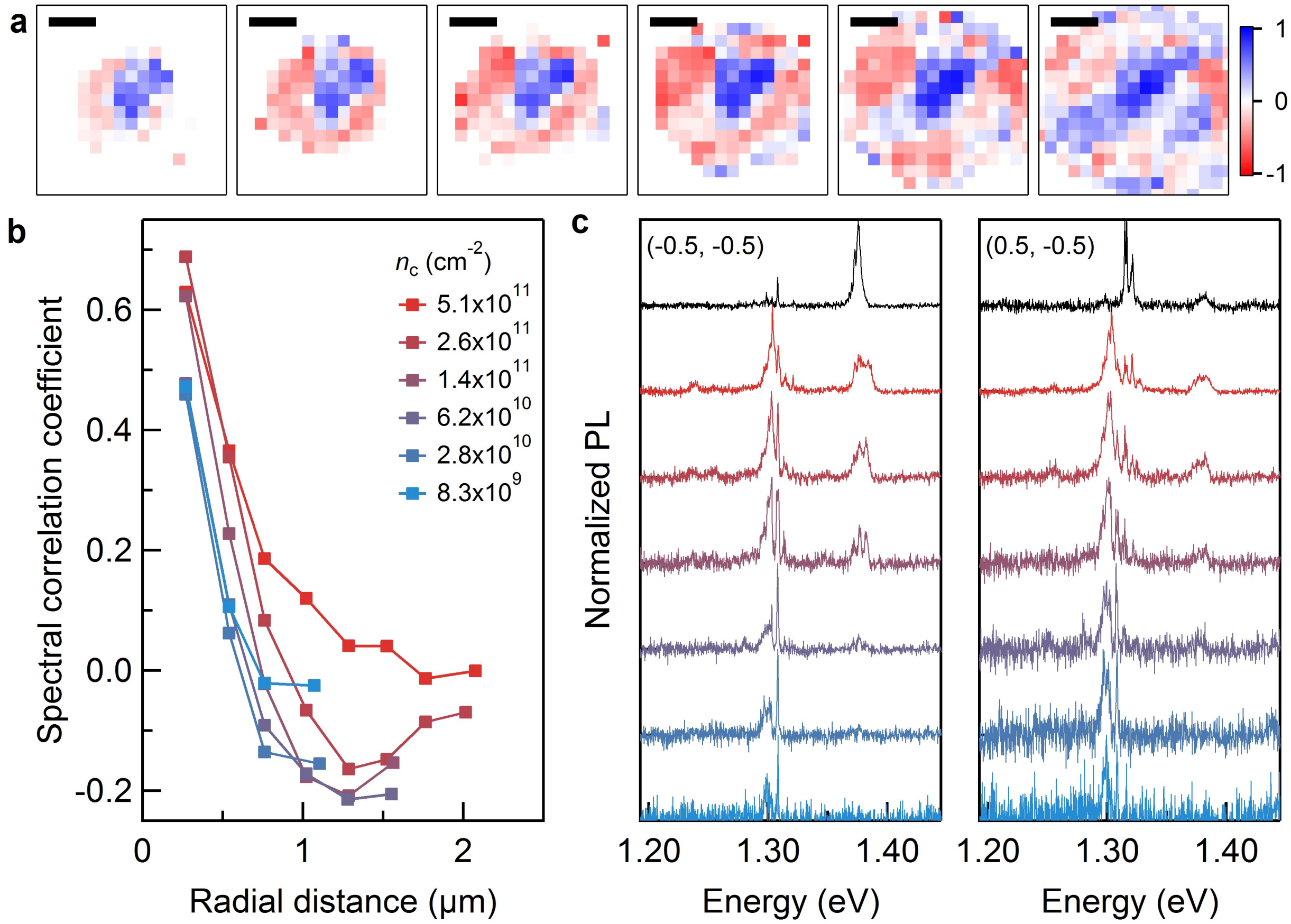

**Fig. S12. Correlation of PL spectra in CW diffusion experiments.** (**a**) Image of PL spectral correlation coefficient between CW diffusion (Fig. 3c) and mapping (Fig. S11a). Center excitation densities are $8\times10^{9}$ ~ $5\times10^{11}$ cm-2 from left to right. Scale bar, 1 μm. (**b**) Spectral correlation coefficients between CW diffusion and mapping, as a function of radial distance from excitation spot, at different excitation densities. The correlation increases with increasing density. (**c**) Representative PL of two spots in CW diffusion (red to blue curves, excitation densities same as in **b**) and mapping (black curve). At low densities, the PL spectra in CW diffusion are all similar to the excitation spot; at high densities, the PL spectra in CW diffusion become more similar to mapping data.

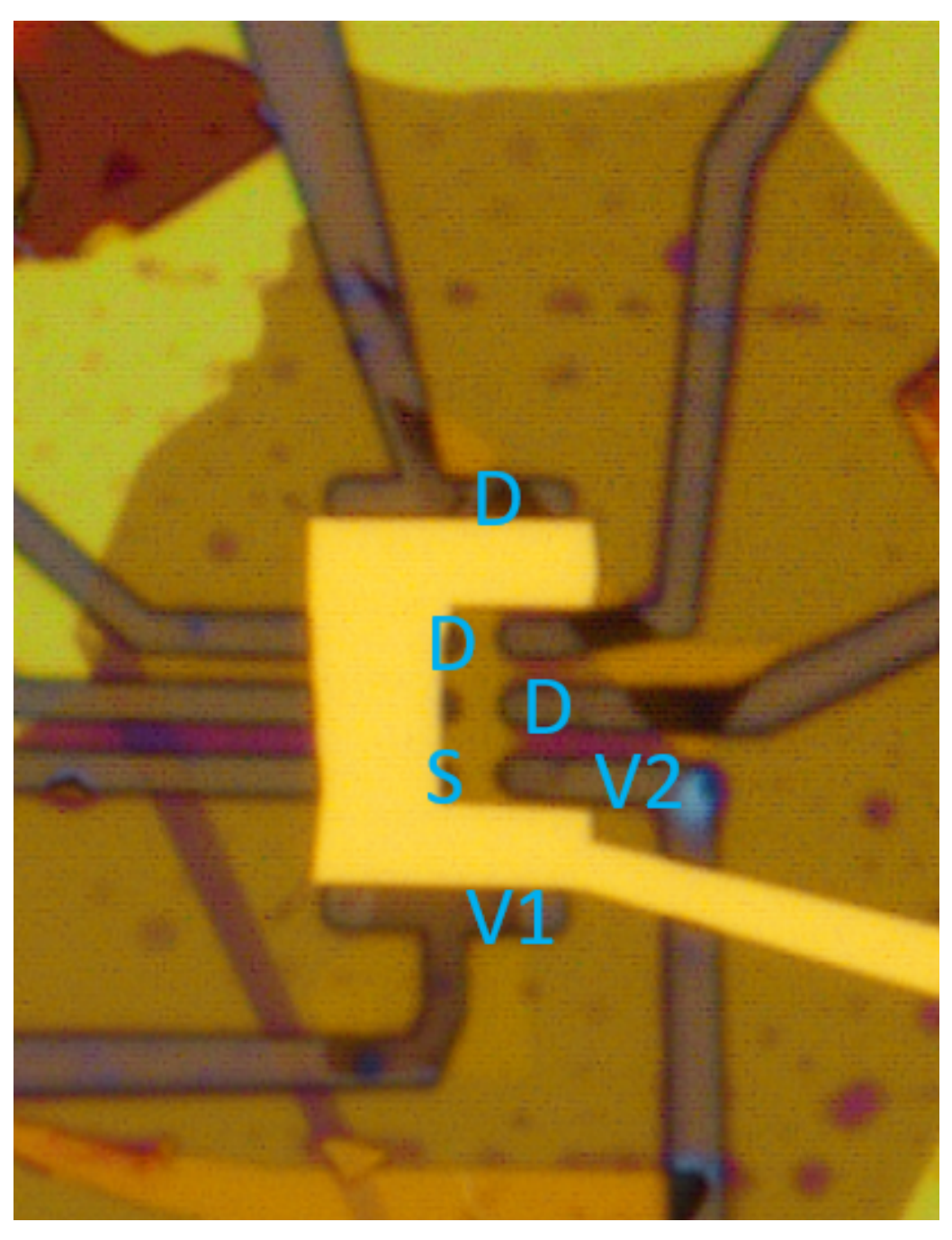

**Fig. S13. The transport device**, with electrodes marked on the optical image. S: source; D: drain; V1 & V2: voltage probes.

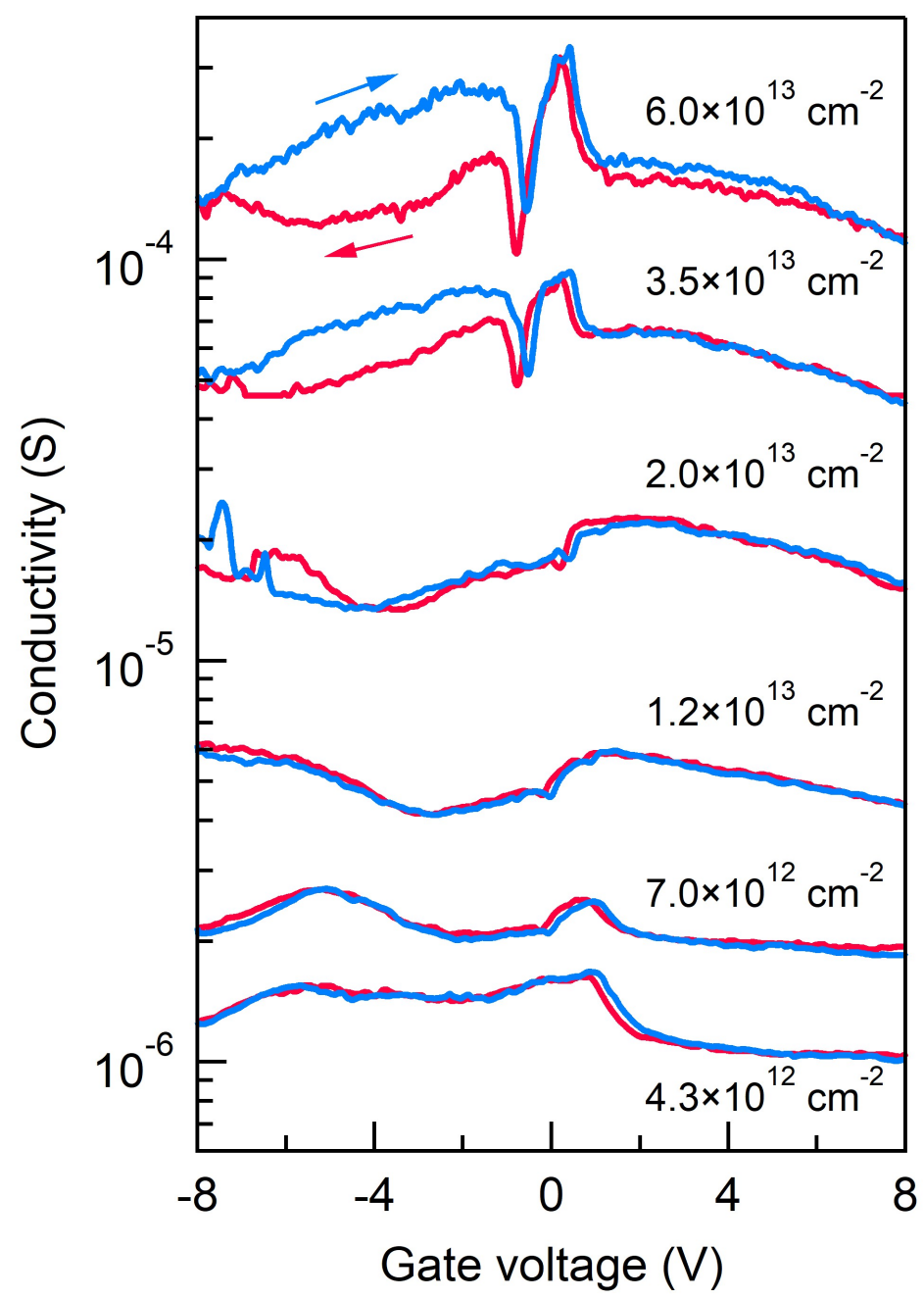

**Fig. S14. Gate voltage dependence of photoconductivity at different steady-state excitation densities.** Gate sweeping with increasing (decreasing) gate voltage is shown as blue (red) curve in each case.

[1] Y. Bai, L. Zhou, J. Wang, W. Wu, L. McGilly, D. Halbertal, C. F. B. Lo, F. Liu, J. Ardelean, P. Rivera, N. R. Finney, X. Yang, D. N. Basov, W. Yao, X. Xu, J. Hone, A. Pasupathy, and X.-Y. Zhu, Nat. Mater. **19**, 10.1038/s41563 (2020).

[2] T. Zhu, J. M. Snaider, L. Yuan, and L. Huang, Annu. Rev. Phys. Chem. **70**, 219 (2019).

[3] N. S. Ginsberg and W. A. Tisdale, Annu. Rev. Phys. Chem. **70**, https://doi.org/10.1146/annurev (2020).

[4] J. Wang, J. Ardelean, Y. Bai, A. Steinhoff, M. Florian, F. Jahnke, X. Xu, M. Kira, J. Hone, and X.-Y. Zhu, Sci. Adv. **5**, eaax0145 (2019).